\documentclass[aps,nofootinbib,superscriptaddress, showpacs,preprintnumbers,  nofootinbibt,twocolumn]{revtex4-2}
\usepackage{eurosym}
\usepackage{amsmath}
\usepackage{bm}
\usepackage{amsfonts}
\usepackage{amssymb}
\usepackage{graphicx}
\usepackage{hyperref}
\usepackage{mathtools}
\usepackage{color}
\usepackage{physics}

\def\be{\begin{equation}}
\def\ee{\end{equation}}
\def\bea{\begin{eqnarray}}
\def\eea{\end{eqnarray}}

\begin{document}

\title{From the Weyl-Schr\"{o}dinger connection to the accelerating Universe - extending Einstein's gravity via a length preserving nonmetricity}
\author{Lei Ming}
\email{minglei@mail.sysu.edu.cn}
\affiliation{ School of Physics,
Sun Yat-Sen University, Guangzhou 510275, People’s Republic of China,}
\author{Shi-Dong Liang}
\email{stslsd@mail.sysu.edu.cn}
\affiliation{ School of Physics,
Sun Yat-Sen University, Guangzhou 510275, People’s Republic of China,}
\author{Hong-Hao Zhang}
\email{zhh98@mail.sysu.edu.cn}
\affiliation{ School of Physics,
Sun Yat-Sen University, Guangzhou 510275, People’s Republic of China,}
\author{Tiberiu Harko}
\email{tiberiu.harko@aira.astro.ro}
\affiliation{Department of Theoretical Physics, National Institute of Physics
and Nuclear Engineering (IFIN-HH), Bucharest, 077125 Romania,}
\affiliation{Department of Physics, Babes-Bolyai University, Kogalniceanu Street,
	Cluj-Napoca, 400084, Romania,}
\affiliation{Astronomical Observatory, 19 Ciresilor Street,
	Cluj-Napoca 400487, Romania}

\begin{abstract}
One of the important extensions of Riemann geometry is Weyl geometry, which is essentially based on the ideas of conformal invariance and nonmetricity. A similar non-Riemannian geometry was proposed by Erwin Schr\"{o}dinger in the late 1940s, in a geometry which is simpler, and (probably) more elegant than the Weyl geometry.  Even it contains  nonmetricity, the Schr\"{o}dinger connection preserves the length of vectors under parallel transport, and thus seems to be more physical than the Weyl connection. Interestingly enough, Schr\"{o}dinger's approach did not attract much interest in the field of gravitational physics. It is the goal of the present paper to reconsider the Schr\"{o}dinger geometry as a potential candidate for a gravitational theory extending standard general relativity. We consider a gravitational action constructed from a length preserving non-metricity, in the absence of torsion, and investigate its variation in both Palatini and metric formalisms. While the Palatini variation leads to standard general relativity, the metric version of the theory adds some non-metricity dependent extra terms in the gravitational Einstein equations, which can be interpreted as representing a geometric type dark energy. After obtaining the generalized Friedmann equations, we analyze in detail the cosmological implications of the theory, by considering two distinct models, corresponding to a dark energy satisfying a linear equation of state, and to conserved matter energy, respectively. In both cases we compare the predictions of the Weyl-Schr\"{o}dinger cosmology with a set of observational data for the Hubble function, and with the results of the $\Lambda$CDM standard paradigm. Our results show that the Weyl-Schr\"{o}dinger cosmological models give a good description of the observational data, and, for certain values of the model parameters, they can reproduce almost exactly the predictions of the $\Lambda$CDM model. Hence, the Weyl-Schr\"{o}dinger theory represents a simple, and viable alternative to standard general relativity, in which dark energy is of purely geometric origin.
\end{abstract}

\date{\today }
\maketitle
\tableofcontents


\section{Introduction}

The creation of the theory of general relativity, as realized in the essential contributions by Einstein and Hilbert \cite{Ein1,Hilbert,Ein2} had an overwhelming impact not only on the various branches of the gravitational physics, including cosmology, but also on mathematics. In their theoretical approaches, Einstein and Hilbert extensively applied the Riemannian geometry \cite{Riemm}, in which on a manifold one can introduce an additional structure, the metric, determined by a metric tensor $g_{\mu \nu}$, and a symmetric connection $\Gamma ^{\alpha}_{\mu \nu}$, respectively. The metric tensor allows us to define distances and angles, while with the help of the connection one can define the covariant derivative $\nabla _\lambda$ of a vector $V_\mu$ as $\nabla _\lambda V_\mu =\partial _\lambda V_\mu-\Gamma ^\sigma _{\lambda \mu}V_\sigma$. The geometric properties of the space time manifold are characterized  by the curvature tensor $R^{\mu }_{\nu \sigma \lambda }$, constructed from the connection, and its contractions, from which the Einstein tensor $G_{\mu \nu}$ is obtained.

In 1918, a few years after the birth of general relativity, Weyl \cite{Weyl, WeylBook} did propose a generalization of Riemannian geometry, which was inspired by the idea of developing the first unified theory of gravity and electromagnetism. In generalizing Riemann geometry, Weyl abandoned the metric condition $\nabla _\lambda g_{\mu \nu}=0$, by generalizing it to $\nabla _\lambda g_{\mu \nu}=Q_{\lambda \mu \nu}$, where $Q_{\lambda \mu \nu}$ is the nonmetricity of the spacetime. In the initial formulation by Weyl,  the nonmetricity has the form $Q_{\lambda \mu \nu}=\omega _\lambda g_{\mu \nu}$, where $\omega _\lambda$ is the Weyl vector. Weyl suggested that the nonmetricity of the spacetime is the source of the electromagnetic field. Weyl's unified theory was severely criticized by Einstein, leading essentially to its abandonment for more than a half century.
Einstein's criticism can be summarized as follows. Under a rescaling of the metric tensor $g_{\mu \nu}\rightarrow \left(1+\epsilon \omega\right)g_{\mu \nu}$, the line element $ds^2 = g_{\mu \nu}dx^\mu dx^\nu$
is rescaled according to $ds\rightarrow  \exp^ {\left(\omega /2\right)}ds$. Einstein asserted that since $ds$ represents the ticking of a clock, or the spacings of atomic spectral lines,  if it is not absolutely invariant, the basic physical quantities (Compton wavelength, electron mass, etc.) would vary in space and time, an effect which is not observed experimentally. This pathological behavior is called the second clock effect. For a recent discussion of it see \cite{Sec}
For early discussions of the Weyl geometry and of its applications see \cite{EddBook} and \cite{PauliBook}, respectively.

Soon after the publication of Weyl's work, another fundamental advance occurred in differential geometry, namely, the definition of the concept of torsion  \cite{Car1}. Theories based on torsion represent another interesting generalization of Einstein's general relativity \cite{Car2,Car3,Car4}, presently called the Einstein-Cartan theory \cite{Hehl1}. In the Einstein-Cartan theory, the torsion field $T^{\mu }_{\sigma \lambda }\neq 0$ is assumed to be proportional to the spin density of the matter \cite{Hehl1}.

It is also worth mentioning, for the sake of completeness,  a third  mathematical and physical enlargement of the gravitational field theories. This extension was initiated  by the work of Weitzenb\"{o}ck \cite{Weitz}, who introduced a class of new  geometrical structures,  known as the Weitzenb\"{o}ck spaces. A Weitzenb\"{o}ck space is characterized by the basic mathematical properties $\nabla _{\mu }g_{\sigma \lambda }= 0$, $T^{\mu }_{\sigma \lambda }\neq 0$, and $R^{\mu }_{\nu \sigma \lambda }=0$, respectively,  and  when $T^{\mu }_{\sigma \lambda }= 0$, reduces  to a Euclidean manifold. Moreover, in a Weitzenb\"{o}ck manifold $T^{\mu }_{\sigma \lambda }$ has values that depend on the regions of  the manifold. Due to the fact that the Riemann curvature tensor identically vanishes,  the  Weitzenb\"{o}ck geometries have the property of distant parallelism, known also as absolute parallelism teleparallelism.  Einstein was the first to apply teleparallelism in physics by proposing a unified teleparallel theory of electromagnetism and gravitation  \cite{Ein}. Weitzenb\"{o}ck geometries are extensively used in  Teleparallel Equivalent of General Relativity (TEGR) type theories, proposed initially in \cite{TE1,TE2,TE3}, also known as the $f(\mathbb{T})$ gravity theory, where $\mathbb{T}$ is the torsion scalar. These theories can explain the late-time acceleration of the Universe, without introducing the dark energy, or the cosmological constant \cite{TE20,TE21,TE22, TE23}. For a review of teleparallel gravity see \cite{revs}.

 With a few notable exceptions, in the physics community Weyl's geometry was almost totally ignored in the first 50 years of its existence. But this situation began to change especially after 1970, when the interest for the physical and mathematical applications of Weyl geometry at both macroscopic and microscopic levels significantly increased. For a detailed description of the fascinating history of Weyl geometry, and of its applications in physics see \cite{Scholz}.

  An important development related to Weyl geometry can be related to the investigations by Dirac \cite{Dirac1,Dirac2}. In proposing an extension of Weyl's theory, and geometry, Dirac introduced the Lagrangian
 \be\label{Dirac}
 L=-\beta ^2R+kD^{\mu}\beta D_{\mu}\beta +c\beta ^4+\frac{1}{4}F_{\mu \nu}F^{\mu \nu},
 \ee
  which contains a real scalar field $\beta$ of weight $w(\beta)=-1$, and the electromagnetic field tensor $F_{\mu \nu}$ coming from the Weyl curvature. Moreover, Dirac adopted for the constant $k$ the value $k=6$. The Lagrangian (\ref{Dirac}) has the important property of  conformal invariance. In \cite{Rosen} the cosmological implications of a Dirac type model were investigated. The Weyl-Dirac type Lagrangian
 \bea
 L&=&W^{\lambda \rho}W_{\lambda \rho}-\beta ^2R+\sigma \beta ^2w^{\lambda}w_{\lambda}+2\sigma \beta w^{\lambda}\beta _{,\lambda}+\nonumber\\
&& (\sigma +6)\beta _{,\rho}\beta_{,\lambda }g^{\rho \lambda}+2\Lambda \beta ^4+L_m,
 \eea
was considered in \cite{Isrcosm},  where $\beta $ is the Dirac scalar field, while $\sigma$ and $\Lambda$ are constants. $W_{\mu \nu}$ is the Weyl length curvature tensor, obtained from the Weyl vector $w_{\mu}$. In the cosmological applications of this model it was shown that ordinary matter is created by the Dirac’s gauge function in the very early Universe. On the other hand, at late times, Dirac’s gauge function generates to dark energy that accelerates the present day Universe..

 Weyl's geometry can be  naturally generalized to include torsion, thus leading to the Weyl-Cartan geometry,which was intensively studied from both mathematical and physical points of view \cite{WC1,WC2,WC3,WC4,WC8, WC9}. For the physical applications of the Riemann-Cartan and Weyl-Cartan space-times  see the review \cite{Rev}.
 A class of teleparallel gravity models, called Weyl-Cartan-Weitzenb\"{o}ck gravity, was proposed in \cite{WCW}, with the action formulated with the help of the dynamical variables $\left(g_{\mu \nu }, w_{\mu
},T^{\lambda}_{~\mu\nu}\right)$.  The teleparallel gravity and the Weyl-Cartan-Weitzenb\"{o}ck theory was generalized in \cite{WCW1}, by inserting the Weitzenb\"{o}ck condition into the Weyl-Cartan gravitational action via a Lagrange multiplier. The cosmological analysis of the theory shows that both accelerating and decelerating cosmological models can be obtained.

 The theoretical investigations performed by using Riemannian, Cartan and teleparallel geometries indicate that general relativity, or more generally, geometric theories of gravity,  can be formulated  in (at least) two formalisms, which are mathematically equivalent: the curvature representation (with the nonmetricity and torsion vanishing identically), and the teleparallel representation, in which the nonmetricity and the curvature vanish identically.

 A third, mathematically equivalent geometric representation of general relativity has also been formulated. The properties of the gravitational field can be described geometrically by the nonmetricity $Q$ of the metric. From a geometric point of view  the nonmetricity describes the change of the length of a vector when parallelly transported around a closed loop. The gravitational theory describing gravity via nonmetricity is called the symmetric teleparallel theory, and it was initially introduced in \cite{Nester}. The connection describing the geometry can be decomposed generally into the Levi-Civita connection of the Riemannian geometry, and a deformation one form, $A^{\alpha}_{\;\;\beta}$, so that $\Gamma ^{\alpha}_{\;\;\beta}=\Gamma ^{\{\}\alpha}_{\;\;\;\;\;\beta}-A^{\alpha}_{\;\;\beta}$. The deformation one form is generally given by  $A_{\alpha \beta}=K_{\alpha \beta}-Q_{\alpha \beta}/2-Q_{\gamma [\alpha \beta]}\theta ^{\gamma}$, where $K_{\alpha \beta}$ is the contorsion, while $Q_{\alpha \beta}$ denotes the nonmetricity, which is generally defined according to $Q_{\alpha \beta}=-Dg_{\alpha \beta}$.

  In a teleparallel frame, in which the condition $\Gamma \equiv 0$ is satisfied, and  after also requiring the condition of the vanishing of the torsion, it follows that $Q_{\mu \nu \lambda}=-g_{\mu \nu,\lambda}$. Hence,  the deformation tensor becomes the Christoffel symbol $\gamma ^{\alpha}_{\beta \gamma}$, so that $A^{\alpha}_{\;\;\beta \gamma}=\gamma ^{\alpha}_{\beta \gamma}$. Then the gravitational action can be represented as $L_g=\sqrt{-g}g^{\mu \nu}\left(\gamma ^{\alpha}_{\beta \mu}\gamma ^{\beta}_{\nu \alpha}-\gamma ^{\alpha}_{\beta \alpha}\gamma ^{\beta}_{\mu \nu }\right)$, which is exactly the Einstein-Hilbert Lagrangian. In symmetric teleparallel gravity the associated energy-momentum density is the Einstein pseudotensor, which now becomes a true tensor.

  The symmetric teleparallel gravity approach was generalized to the $f(Q)$ gravity theory (or the coincident general relativity) in \cite{Lav}. As a first step in constructing the theory one introduces the quadratic nonmetricity scalar
  \be
  Q=-\frac{1}{4}Q_{\alpha \beta \mu}Q^{\alpha \beta \mu}+\frac{1}{2}Q_{\alpha \beta \mu}Q^{\beta \mu \alpha}+\frac{1}{4}Q_{\alpha}Q^{\alpha}-\frac{1}{2}Q_{\alpha}\bar{Q}^{\alpha},
  \ee
  where $Q_{\mu}=Q^{\;\;\alpha}_{\mu \;\;\alpha}$, and $\tilde{Q}^{\mu}=Q_{\alpha}^{\;\;\mu \alpha}$. Then, the nonmetricity conjugate $P^{\alpha}_{ \;\;\mu \nu}$ is defined as
  \bea
  P^{\alpha}_{ \;\;\mu \nu}&=&c_1Q^{\alpha \;\mu \nu}+c_2Q^{\;\;\alpha}_{(\mu \;\nu)}+c_3Q^{\alpha}g_{\mu\nu}+c_4\delta^{\alpha}_{(\mu}\bar{Q}_{\nu)}\nonumber\\
  &&+\frac{c_5}{2}\left(\tilde{Q}^{\alpha}g_{\mu \nu}+\delta^{\alpha}_{(\mu }Q_{\nu)}\right).
  \eea
  Finally, after introducing the general quadratic form $\mathrm{Q}$ as $\mathrm{Q}=Q_{\alpha}^{\;\mu \nu}P^{\alpha}_{\;\mu \nu}$, one can write down the gravitational action of the $f(Q)$ theory as \cite{Lav}
  \be
  S=\int{d^nx\left[-\frac{1}{2}\sqrt{-g}\mathrm{Q}+\lambda _{\alpha}^{\;\beta \mu \nu}R^{\alpha}_{\;\beta \mu \nu}+\lambda _{\alpha}^{\;\mu \nu}T^{\alpha}_{\;\mu \nu}\right]}.
  \ee

  The physical, cosmological and geometrical properties of the $f(Q)$ gravity have been extensively investigated recently \cite{Q1,Q2,Q3,Q4,Q5,Q6,Q7}. For a review of the $f(Q)$ theory see \cite{revQ}.  An extension of the $f(Q)$ theory was considered in \cite{s12} by assuming that the nonmetricity $Q$ could nonminimally couple to the matter Lagrangian. The action of the theory is thus given by
\be
S=\int{d^4x\sqrt{-g}\left[\frac{1}{2}f_1(Q)+f_2(Q)L_m\right]},
\ee
where $L_m$ is the matter Lagrangian, and $f_1$ and $f_2$ are arbitrary analytical functions of $Q$. The existence of a nonminimal coupling between geometry and matter leads to the nonconservation of the matter energy-momentum tensor, and to the presence of an extra force in the geodesic equation of motion of massive particles. The cosmological solutions obtained in the framework of this model can describe the  accelerating evolution of the Universe.

 The most general extension of the $f(Q)$ gravity, with the gravitational Lagrangian $L$ constructed  from an arbitrary function $f$ of the non-metricity $Q$ and of the trace $T$ of the matter-energy-momentum tensor, was introduced in \cite{s21}. The action of the theory is
 \be
 S=\int{\left[\frac{1}{16\pi}f(Q,T)+L_m\right]\sqrt{-g}d^4x},
 \ee
 Within the framework of $f(Q,T)$ gravity one can construct cosmological models by assuming some simple functional forms of the function $f( Q, T)$. In these models the Universe enters in an accelerating phase, which usually ends with a de Sitter type expansion.

 In the 1940's Erwin Schr\"{o}dinger, who was mostly interested in metric-affine theories, tried to find the most general possible symmetric connection  \cite{Sch1, SchroBook}.  From general considerations he arrived at the result that such a connection is given by $\Gamma ^\lambda _{\mu \nu}=\gamma _{\mu \nu}^\lambda+g^{\lambda \rho}S_{\rho \mu \nu}$, where $S_{\rho \mu \nu}$ is a geometric quantity consisting of the combination of two antisymmetric connections. The geometry based on Schr\"{o}dinger's connection represents essentially a new geometry, which is distinct from that of Weyl. A systematic investigation of the Schr\"{o}dinger connection was performed in
\cite{Klemm:2020gfm}, where an action of the form
\begin{equation}\label{actiontot}
\begin{split}
S & = \frac1{2\kappa^2}\int d^3 x\left(\sqrt{-g} f(R) + \frac1{2\mu} \epsilon^{\mu\nu\rho} Q_\rho
\hat{R}_{\nu\mu}\right) \\
& + \int d^3 x\,\epsilon^{\mu\nu\rho}\zeta_{\nu\sigma} {T_{\rho\mu}}^\sigma\,,
\end{split}
\end{equation}
was considered, where  $f(R)$ denotes an arbitrary function of the scalar curvature $R = g^{\mu\nu} R_{\mu\nu}(\Gamma)$, $\hat{R}_{\mu\nu} := {R^\lambda}_{\lambda\mu\nu} =
\partial_{[\mu}Q_{\nu]}$ is the homothetic curvature tensor, $\mu$ is a Chern-Simons coupling constant, and $\zeta_{\nu\sigma}$ a Lagrange multiplier, respectively. Moreover,
$\epsilon^{\mu\nu\rho}=\sqrt{-g}\varepsilon^{\mu\nu\rho}$ the Levi-Civita symbol,  $\varepsilon^{\mu\nu\rho}$ is the Levi-Civita tensor, while $\Gamma $ denotes an arbitrary connection.  Solutions with constant scalar curvature were obtained in the framework of this model, leading to a self-duality relation for the nonmetricity vector. This relation gives a Proca type equation, which may be an indication of the inhomogeneous Maxwell equations as originating from affine geometry.

\begin{figure}[htbp]
	\includegraphics[scale=0.1]{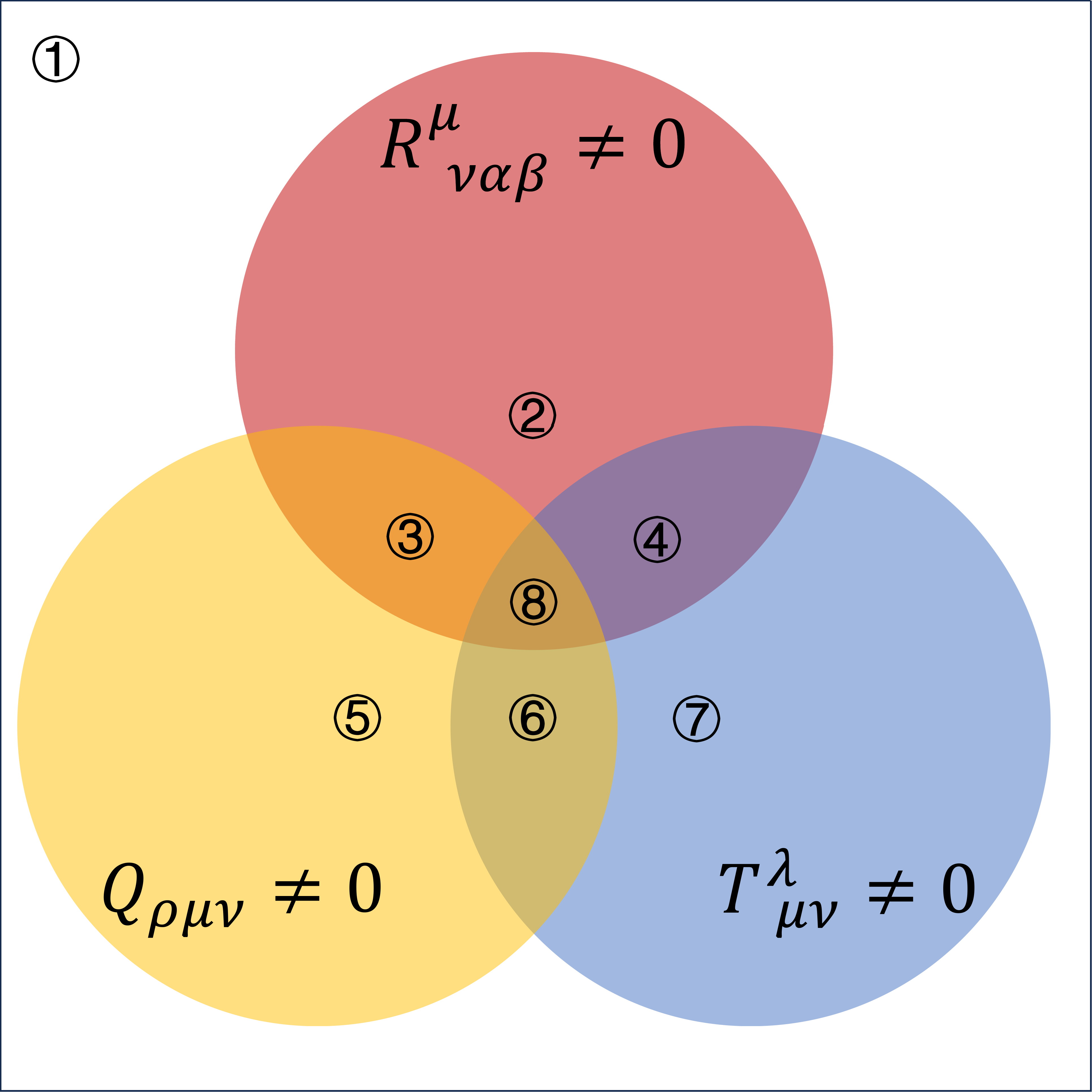}
	\caption{Eight possibilities of gravity theories (geometries): (1) Minkowski, (2) Riemann, (3) Weyl and Schr\"{o}dinger, (4) Cartan, (5) symmetric teleparallel, (6) generic teleparallel, (7) metric teleparallel(Weitzenb\"{o}ck) and (8) metric-affine.}
	\label{fig0}
\end{figure}

We have seen that a general geometry of spacetime can be characterized by three geometric variables: curvature, torsion and non-metricity. Thus, in different theories of gravity one could decide whether to include any of them. This gives us eight possible choices, as shown in Fig.~\ref{fig0}. The most general gravity theory that contains all of the three variables is based on a non-Riemannian geometry and is called Metric-Affine gravity, while the most trivial case with all three being zero leads to a Minkowski spacetime. The Schr\"{o}dinger geometry that we are going to study in detail and the Weyl geometry belong to the same category, in which both curvature and non-metricity are presented.

It is the main goal of the present investigation to consider the possibility of the Schr\"{o}dinger geometry as being an important and viable candidate for the geometric extension of standard general relativity. To implement this idea we begin by considering a gravitational action, which is formulated in terms of a length preserving non-metricity, in the absence of torsion.  The variation of this action is considered in both Palatini and metric formalisms. It turns out that the Palatini variation leads to standard general relativity, and hence the two theories coincide in this formulation. However, the metric variation of the Schr\"{o}dinger action leads to the presence of  non-metricity dependent extra terms in the gravitational Einstein equations.  We interpret these terms as representing a geometric type dark energy.

In order to investigate the physical implications, and the viability of the Weyl-Schr\"{o}dinger theory we consider the field equations in the FLRW cosmological metric. After deriving the generalized Friedmann equations, we analyze in detail the cosmological implications of the Weyl-Schr\"{o}dinger theory. In the generalized Friedmann equations the presence of nonmetricity generates two new terms, which can be interpreted as an effective geometric energy density of the dark matter, and an effective pressure.  To test viability of the theory we consider two distinct cosmological models. In the first model, we assume that  dark energy satisfies a linear equation of state, that is, the effective geometric pressure is proportional to the dark matter energy density, with the parameter of the equation of state assumed to be a redshift dependent function. In the second model
we  assume that the matter energy density is conserved. For both models we perform a comparison of the predictions of the Weyl-Schr\"{o}dinger cosmology with a set of observational data for the Hubble function, and with the similar results obtained within the framework of the $\Lambda$CDM standard cosmological paradigm. Our results show that the Weyl-Schr\"{o}dinger cosmological models can give a good description of the observational data for the Hubble function. Moreover, for specific values of the model parameters, they can reproduce almost exactly the predictions of the $\Lambda$CDM model. Therefore, the Weyl-Schr\"{o}dinger theory in its nonmetricity representation could provide a simple, and viable alternative to standard general relativity, in which dark energy is of purely geometric origin.

The present paper is organized as follows. We introduce the fundamentals of the Weyl and Schr\"{o}dinger geometries in Section~\ref{sect1}. The action of the Weyl-Schr\"{o}dinger theory is introduced in Section~\ref{sect2}, where the gravitational field equations are derived in both Palatini and metric formalisms.  The cosmological implications of the theory are investigated in Section~\ref{sect3}, where the predictions of two distinct cosmological models are compared with the observational data, and the similar predictions of the $\Lambda$CDM model.  A thermodynamic interpretation of the Weyl-Schr\"{o}dinger gravity is presented in Section~\ref{sect4}. We discuss our results, and we conclude our work in Section~\ref{sect5}. The calculational details of the variation of the action in the Palatini formalism are presented in Appendix~\ref{App_var}. The technical details of the calculation of the variation of the action with respect to the metric tensor are given in Appendix~\ref{varg}. Finally, the derivation of the generalized Friedmann equations for the FLRW metric is presented in Appendix~\ref{App_Fri}.

\section{From Weyl geometry to the Schr\"{o}dinger connection}\label{sect1}

In his book \cite{SchroBook}, Schr\"{o}dinger wished to find the most general class of an affine connection to be in accordance with the affine measure of distance along every geodesic, i.e., a relationship between $g_{\mu\nu}$ and $\Gamma^{\lambda}_{~\mu\nu}$. While allowing the existence of a non-zero nonmetricity, a Schr\"{o}dinger connection is supposed to preserve the length of vectors under parallel transport, which in general does not hold in Weyl geometry. To find such a form of connection, one can start with a sufficient condition with vanishing nonmetricity that $Q_{\alpha\mu\nu}=-\nabla_\alpha g_{\mu\nu}=0$, then the circling of it gives
\bea
0&=&\nabla_\rho g_{\mu\nu}=\partial_\rho g_{\mu\nu}-g_{\mu\alpha}\Gamma^{\alpha}_{~\nu\rho}-g_{\nu\alpha}\Gamma^{\alpha}_{~\mu\rho},\\
0&=&\nabla_\mu g_{\nu\rho}=\partial_\mu g_{\nu\rho}-g_{\nu\alpha}\Gamma^{\alpha}_{~\rho\mu}-g_{\rho\alpha}\Gamma^{\alpha}_{~\nu\mu},\\
0&=&\nabla_\nu g_{\rho\mu}=\partial_\nu g_{\rho\mu}-g_{\rho\alpha}\Gamma^{\alpha}_{~\mu\nu}-g_{\mu\alpha}\Gamma^{\alpha}_{~\rho\nu}.
\eea
Adding the later two equations and minus the first, and contracting with $\frac{1}{2}g^{\rho\lambda}$ yields
\begin{align}
0&=\frac{1}{2}g^{\rho\lambda}\left(\partial_\mu g_{\nu\rho}+\partial_\nu g_{\rho\mu}-\partial_\rho g_{\mu\nu}\right)-\frac{1}{2}g^{\rho\lambda}g_{\rho\alpha}\left(\Gamma^\alpha_{~\mu\nu}+\Gamma^{\alpha}_{~\nu\mu}\right)\nonumber\\
&+\frac{1}{2}g^{\rho\lambda}g_{\nu\alpha}\left(\Gamma^\alpha_{~\mu\rho}-\Gamma^{\alpha}_{~\rho\mu}\right)+\frac{1}{2}g^{\rho\lambda}g_{\mu\alpha}\left(\Gamma^\alpha_{~\nu\rho}-\Gamma^{\alpha}_{~\rho\nu}\right)\nonumber\\
&=\gamma^\lambda_{\mu\nu}-\Gamma^\lambda_{~(\mu\nu)}+g^{\rho\lambda}\left(g_{\nu\alpha}\Gamma^\alpha_{~[\mu\rho]}+g_{\mu\alpha}\Gamma^\alpha_{~[\nu\rho]}\right),
\end{align}
or
\be\label{orGamma}
\Gamma^\lambda_{~\mu\nu}=\gamma^\lambda_{\mu\nu}+g^{\rho\lambda}\left(g_{\nu\alpha}\Gamma^\alpha_{~[\mu\rho]}+g_{\mu\alpha}\Gamma^\alpha_{~[\nu\rho]}\right)+\Gamma^\lambda_{~[\mu\nu]},
\ee
where $\gamma^\lambda_{\mu\nu}=\frac{1}{2}g^{\rho\lambda}\left(\partial_\mu g_{\nu\rho}+\partial_\nu g_{\rho\mu}-\partial_\rho g_{\mu\nu}\right)$ is the Christoffel symbol, $\Gamma^\lambda_{~(\mu\nu)}\equiv\frac{1}{2}\left(\Gamma^\lambda_{\mu\nu}+\Gamma^\lambda_{~\nu\mu}\right)$ and $\Gamma^\lambda_{~[\mu\nu]}\equiv\frac{1}{2}\left(\Gamma^\lambda_{\mu\nu}-\Gamma^\lambda_{~\nu\mu}\right)$ are the symmetric and antisymmetric parts of $\Gamma^\lambda_{~\mu\nu}$, respectively. Considering that in the equations of geodesic, antisymmetry cancels in the overall connection, we can drop the last term in (\ref{orGamma}) and are led to
\be
\Gamma^\lambda_{~\mu\nu}=\gamma^\lambda_{\mu\nu}+g^{\rho\lambda}\left(g_{\nu\alpha}\Gamma^\alpha_{~[\mu\rho]}+g_{\mu\alpha}\Gamma^\alpha_{~[\nu\rho]}\right).
\ee

From this  relation Schr\"{o}dinger concluded that the connections $\Gamma^\lambda_{~\mu\nu}$, which are compatible with the metric $g_{\mu\nu}$ in a general meaning, while the condition $Q_{\rho\mu\nu}=0$ is not necessary satisfied, should have the form
\be\label{S_con}
\Gamma^\lambda_{~\mu\nu}=\gamma^\lambda_{\mu\nu}+g^{\rho\lambda}S_{\rho\mu\nu},
\ee 	
where the Schr\"{o}dinger tensor $S_{\rho\mu\nu}$ is symmetric in its later indices
\be \label{S_sym}
S_{\rho\mu\nu}=S_{\rho\nu\mu}.
\ee
He also concluded that with this form of connections, the necessary condition that a vector preserves its length under parallel transport is
\be\label{S_cir}
S_{(\rho\mu\nu)}=0.
\ee
To see this, notice that for a tangent vector of a geodesic, $\xi^\mu\propto\frac{dx^\mu}{d\lambda}$, we have $\xi^\rho\nabla_\rho \xi^\mu=0$, then its length being constant gives
\begin{align}
	0=&\xi^\rho\nabla_\rho(g_{\mu\nu}\xi^\mu\xi^\nu)\nonumber\\
	=&\nabla_\rho g_{\mu\nu}\xi^\rho\xi^\mu\xi^\nu+g_{\mu\nu}\xi^\nu\xi^\rho\nabla_\rho\xi^\mu+g_{\mu\nu}\xi^\mu\xi^\rho\nabla_\rho\xi^\nu\nonumber\\
	=&\left(\partial_\rho g_{\mu\nu}-g_{\mu\alpha}\Gamma^\alpha_{~\nu\rho}-g_{\nu\alpha}\Gamma^\alpha_{~\mu\rho}\right)\xi^\rho\xi^\mu\xi^\nu\nonumber\\
	=&\left(\partial_\rho g_{\mu\nu}-g_{\mu\alpha}\gamma^\alpha_{\nu\rho}-g_{\nu\alpha}\gamma^\alpha_{\mu\rho}-g_{\mu\alpha}g^{\lambda\alpha}S_{\lambda\nu\rho}\right.\nonumber\\
	&\left.-g_{\nu\alpha}g^{\lambda\alpha}S_{\lambda\mu\rho}\right)\xi^\rho\xi^\mu\xi^\nu\nonumber\\
	=&-\left(S_{\mu\nu\rho}+S_{\nu\mu\rho}\right)\xi^\rho\xi^\mu\xi^\nu\nonumber\\
	=&-2S_{\mu\nu\rho}\xi^\rho\xi^\mu\xi^\nu,
\end{align}
and thus we arrive at $S_{(\mu\nu\rho)}=0$.

To summarize, a Schr\"{o}dinger connection, which preserves the length of vectors under parallel transport, although involving non-zero nonmetricity, has the form (\ref{S_con}), while fulfilling the conditions (\ref{S_sym}) and (\ref{S_cir}), respectively.

As is well known, the generic decomposition of an affine connection is given by
\begin{align}\label{Gamma_dec}
	\Gamma^\lambda_{~\mu\nu}=&\gamma^\lambda_{~\mu\nu}+N^\lambda_{~\mu\nu}\nonumber\\
	=&\gamma^\lambda_{~\mu\nu}+L^\lambda_{~\mu\nu}+C^\lambda_{~\mu\nu}\nonumber\\
	=&\gamma^\lambda_{~\mu\nu}+\frac{1}{2}g^{\rho\lambda}\left(Q_{\mu\nu\rho}+Q_{\nu\rho\mu}-Q_{\rho\mu\nu}\right)\nonumber\\
	&+g^{\rho\lambda}\left(T_{\rho\mu\nu}+T_{\mu\nu\rho}-T_{\nu\rho\mu}\right),
\end{align}
where $T^\lambda_{~\mu\nu}\coloneqq\Gamma^\lambda_{~[\mu\nu]}$ is the torsion tensor, $N^\lambda_{~\mu\nu}$, $L^\lambda_{~\mu\nu}$ and $C^\lambda_{~\mu\nu}$ are the distortion, deflection and contorsion tensor, respectively. In the case of symmetric connection (i.e., vanishing torsion) and $N_{(\rho\mu\nu)}=0$, (\ref{Gamma_dec}) reduces to
\begin{align}\label{S_Gamma}
	\Gamma^\lambda_{~\mu\nu}=&\gamma^\lambda_{\mu\nu}+\frac{1}{2}g^{\rho\lambda}\left(Q_{\mu\nu\rho}+Q_{\nu\rho\mu}-Q_{\rho\mu\nu}\right)\nonumber\\
	=&\gamma^\lambda_{~\mu\nu}+\frac{1}{2}g^{\rho\lambda}\left(-Q_{\rho\mu\nu}-Q_{\rho\mu\nu}\right)\nonumber\\
	=&\gamma^\lambda_{\mu\nu}-g^{\rho\lambda}Q_{\rho\mu\nu}
\end{align}
with $Q_{(\rho\mu\nu)}=0$. This corresponds to a Schr\"{o}dinger connection for which the Schr\"{o}dinger tensor $S_{\rho\mu\nu}=-Q_{\rho\mu\nu}$. It was also discussed that the Schr\"{o}dinger connection can be written only in terms of torsion with vanishing nonmetricity \cite{Klemm:2020gfm}. We will focus on the torsion free case and consider a length preserving nonmetricity.

Having the affine connection \eqref{S_Gamma} at hand, we can then define the Riemann curvature tensor $R^\mu_{~\nu\alpha\beta}$, which describes how parallel transport modifies the orientation of a vector, by acting the commutator of two covariant derivatives on a vector $v^\mu$,
\begin{equation}
[\nabla_\alpha,\nabla_\beta]v^\mu=2\nabla_{[\alpha]}\nabla_{\beta]}v^\mu=R^\mu_{~\nu\alpha\beta}v^\nu,
\end{equation}
where $R^\mu_{~\nu\alpha\beta}$ is related to the Schr\"{o}dinger connection $\Gamma^\lambda_{~\mu\nu}$ via 
\begin{equation}
R^\mu_{~\nu\alpha\beta}\equiv\partial_\alpha\Gamma^\mu_{~\nu\beta}-\partial_\beta\Gamma^\mu_{~\nu\alpha}+\Gamma^\mu_{~\rho\alpha}\Gamma^\rho_{~\nu\beta}-\Gamma^\mu_{~\rho\beta}\Gamma^\rho_{~\nu\alpha}.
\end{equation}
Without the help of metric, there exist two possible independent contractions of $R^\mu_{~\nu\alpha\beta}$, namely the Ricci tensor $R_{\mu\nu}$ and the  homothetic curvature $\hat{R}_{\mu\nu}$,
\begin{equation}
R_{\mu\nu}\equiv R^\alpha_{~\mu\alpha\nu}=\partial_\alpha\Gamma^\alpha_{~\mu\nu}-\partial_\nu\Gamma^\alpha_{~\mu\alpha}+\Gamma^\alpha_{~\rho\alpha}\Gamma^\rho_{~\mu\nu}-\Gamma^\alpha_{~\rho\nu}\Gamma^\rho_{~\mu\alpha}
\end{equation}
and
\begin{equation}
\hat{R}_{\mu\nu}\equiv R^\alpha_{~\alpha\mu\nu}=\partial_\mu\Gamma^\alpha_{~\alpha\nu}-\partial_\nu\Gamma^\alpha_{~\alpha\mu}. 
\end{equation}
It is known that when non-metricity is present ($Q_{\rho\mu\nu}\neq0$), the homothetic curvature $\hat{R}_{\mu\nu}$ is non-vanishing and can be expressed as $\hat{R}_{\mu\nu}=\partial_{[\mu}Q_{\nu]}$. However, in next section we will see that only the symmetric part of Ricci tensor appears in the equations of motion since $\hat{R}_{\mu\nu}$ is anti-symmetric in the indices $\mu$ and $\nu$.  

The Ricci tensor $R_{\mu\nu}$ can be decomposited into two parts, the pure Riemannian part $\mathring{R}_{\mu\nu}$ computed for the Levi-Civita connection $\gamma^\lambda_{\mu \nu}$ and the part containing the contribution of non-metricity, 
\begin{equation}
	\begin{aligned}\label{eq_RR}
		R_{\mu\nu}\equiv&\partial_\alpha\Gamma^\alpha_{~\mu\nu}-\partial_\nu\Gamma^\alpha_{~\mu\alpha}+\Gamma^\alpha_{~\rho\alpha}\Gamma^\rho_{~\mu\nu}-\Gamma^\alpha_{~\rho\nu}\Gamma^\rho_{~\mu\alpha}\\
		=&\partial_\alpha\gamma^\alpha_{\mu\nu}-\partial_\alpha Q^\alpha_{~\mu\nu}-\partial_\nu\gamma^\alpha_{\mu\alpha}+\partial_\nu\tilde{Q}_\mu\\
		&+\gamma^\alpha_{\rho\alpha}\gamma^\rho_{\mu\nu}-\tilde{Q}_\rho\gamma^\rho_{\mu\nu}-\gamma^\alpha_{\rho\alpha}Q^\rho_{~\mu\nu}+\tilde{Q}_\rho Q^\rho_{~\mu\nu}\\
		&-\gamma^\alpha_{\rho\nu}\gamma^\rho_{\mu\alpha}+Q^\alpha_{~\rho\nu}\gamma^\rho_{\mu\alpha}+\gamma^\alpha_{\rho\nu}Q^\rho_{~\mu\alpha}-Q^\alpha_{~\rho\nu} Q^\rho_{~\mu\alpha}\\
		=&\mathring{R}_{\mu\nu}-\mathring{\nabla}_\alpha Q^\alpha_{~\mu\nu}+\mathring{\nabla}_\nu\tilde{Q}_\mu+\tilde{Q}_\rho Q^\rho_{~\mu\nu}-Q^\alpha_{~\rho\nu}Q^\rho_{~\mu\alpha},
	\end{aligned}
\end{equation}
where $\mathring{\nabla}_\alpha$ is the covariant derivative corresponding to $\gamma^\rho_{\mu\nu}$, i.e., $\mathring{\nabla}_\alpha g_{\mu\nu}=0$,
and we use
\bea
&&\mathring{R}_{\mu\nu}\coloneqq\partial_\alpha\gamma^\alpha_{\mu\nu}-\partial_\nu\gamma^\alpha_{\mu\alpha}+\gamma^\alpha_{\rho\alpha}\gamma^\rho_{\mu\nu}-\gamma^\alpha_{\rho\nu}\gamma^\rho_{\mu\alpha},\\
&&\mathring{\nabla}_\alpha Q^\alpha_{~\mu\nu}=\partial_\alpha  Q^\alpha_{~\mu\nu}+\gamma^\alpha_{\rho\alpha}Q^\rho_{~\mu\nu}
-\gamma^\rho_{\mu\alpha}Q^\alpha_{~\rho\nu}-\gamma^\rho_{\nu\alpha}Q^\alpha_{~\mu\rho},\\
&&\mathring{\nabla}_\nu \tilde{Q}_\mu=\partial_\nu\tilde{Q}_\mu-\gamma^\rho_{\mu\nu}\tilde{Q}_\rho.
\eea

Before going further into the Weyl-Schr\"{o}dinger geometry, we shall discuss more about the conditions of the nonmetricity. With the two independent vectors of nonmetricity, $Q_\mu$ and $\tilde{Q}_\mu$, the nonmetricity tensor $Q_{\lambda\mu\nu}$ can be decomposed in $n$ dimension as \cite{Iosifidis:2019dua}
\begin{align}\label{Q_dec}
	&Q_{\lambda\mu\nu}=\frac{n+1}{(n+2)(n-1)}Q_\lambda g_{\mu\nu}-\frac{2}{(n+2)(n-1)}Q_{(\mu} g_{\nu)\lambda}\nonumber\\
	&-\frac{2}{(n+2)(n-1)}\tilde{Q}_\lambda g_{\mu\nu}+\frac{2n}{(n+2)(n-1)}\tilde{Q}_{(\mu}g_{\nu)\lambda}+\Omega_{\lambda\mu\nu},
\end{align}
where $\Omega_{\lambda\mu\nu}$ is the traceless part of $Q_{\lambda\mu\nu}$: $g^{\mu\nu}\Omega_{\lambda\mu\nu}=g^{\lambda\mu}\Omega_{\lambda\mu\nu}=0$.

If we assume $\Omega_{\lambda\mu\nu}=0$ and $\tilde{Q}_\mu=m Q_\mu$ with some constant $m$,
then
\be
Q_{\lambda\mu\nu}=\frac{n+1-2m}{(n+2)(n-1)}Q_\lambda g_{\mu\nu}+\frac{2mn-2}{(n+2)(n-1)}Q_{(\mu} g_{\nu)\lambda},
\ee
thus
\begin{align}
Q_{(\lambda\mu\nu)}=&\frac{n+1-2m}{(n+2)(n-1)}Q_{(\lambda}g_{\mu\nu)}+\frac{2mn-2}{(n+2)(n-1)}Q_{(\mu} g_{\nu\lambda)}\nonumber\\
=&\frac{2m+1}{n+2}Q_{(\lambda}g_{\mu\nu)},
\end{align}
therefore the condition $Q_{(\lambda\mu\nu)}=0$ leads to $m=-\frac{1}{2}$, i.e.,
\be
\tilde{Q}_\mu=-\frac{1}{2}Q_\mu.
\ee
We are going to find a solution of Schr\"{o}dinger connection that satisfies the above condition together with $\Omega_{\lambda\mu\nu}=0$.

\section{Gravitational field equations in the Weyl-Schr\"{o}dinger geometry}\label{sect2}

In this paper we first work in the Palatini formalism in which the affine connection and metric are considered to be two independent variables and  the matter part of the action does not depend on the connection. The gravitational action we will study is
\begin{align}\label{action}
S=\frac{1}{16\pi}\int d^4x& \sqrt{-g}\bigg(R+\frac{5}{24}Q_\rho Q^\rho+\frac{1}{6}\tilde{Q}_\rho\tilde{Q}^\rho+2T_\rho Q^\rho\nonumber\\
&+\zeta^{\rho\sigma}_{~~\alpha}T^\alpha_{~\rho\sigma}\bigg)+\int d^4x\sqrt{-g}L_m,
\end{align}
here $R\coloneqq g^{\mu\nu}R_{\mu\nu}(\Gamma)$, $T_\rho\coloneqq T^\sigma_{~\rho\sigma}$, and $\zeta^{\rho\sigma}_{~~\alpha}$ is a Lagrange multiplier.

The variation of (\ref{action}) with respect to $\zeta^{\mu\nu}_{~~\lambda}$ leads to a vanishing torsion,
\be\label{0T}
T^\lambda_{~\mu\nu}=0.
\ee
The variation w.r.t. the metric $g^{\mu\nu}$ gives the modified Einstein equation,
\begin{align}\label{eq_EinsteinPalatini}
	R_{(\mu\nu)}-\frac{1}{2}Rg_{\mu\nu}+\frac{5}{24}\bigg(\frac{1}{2}g_{\mu\nu}Q_\alpha Q^\alpha
	+Q_\mu Q_\nu-2g_{\mu\nu}Q_\alpha \tilde{Q}^\alpha &\nonumber\\ -2g_{\mu\nu}g^{\alpha\beta}\nabla_\beta Q_\alpha\bigg)+\frac{1}{6}\bigg(-\frac{1}{2}g_{\mu\nu}\tilde{Q}_\alpha\tilde{Q}^\alpha-\tilde{Q}_\mu\tilde{Q}_\nu&\nonumber\\
	+Q_\mu\tilde{Q}_\nu-2\nabla_{(\mu}\tilde{Q}_{\nu)}\bigg)=8\pi T_{\mu\nu},&
\end{align}
where as usual we define the energy-momentum tensor
\be
T_{\mu\nu}\equiv-\frac{2}{\sqrt{-g}}\frac{\delta\left(\sqrt{-g}L_m\right)}{\delta g^{\mu\nu}},
\ee
and varying (\ref{action}) w.r.t $\Gamma^\lambda_{~\mu\nu}$ results in
\begin{align}\label{eq_con}
	-\frac{\nabla_\lambda\left(\sqrt{-g}g^{\mu\nu}\right)}{\sqrt{-g}}+\frac{\nabla_\rho\left(\sqrt{-g}g^{\mu\rho}\right)\delta^\nu_\lambda}{\sqrt{-g}}+\frac{1}{3}\tilde{Q}_\lambda g^{\mu\nu}&\nonumber\\
	+\frac{1}{3}\tilde{Q}^\mu\delta^\nu_\lambda+Q^\mu\delta^\nu_\lambda-\frac{1}{6}Q^\nu\delta^\mu_\lambda+\zeta^{[\mu\nu]}_{~~~\lambda}=0.&
\end{align}
The detailed calculation for these equations of motion can be found in Appendix \ref{App_var}.

Noticing that
\be\label{QL}
\frac{\nabla_\lambda\sqrt{-g}}{\sqrt{-g}}=\frac{1}{2g}\nabla_\lambda g=\frac{1}{2g}g g^{\alpha\beta}\nabla_\lambda g_{\alpha\beta}=-\frac{1}{2}Q_\lambda
\ee
and
\be\label{QLMN}
\hspace{-0.4cm}\nabla_\lambda g^{\mu\nu}=g^{\alpha\mu}g_{\alpha\beta}\nabla_\lambda g^{\beta\nu}=-g^{\alpha\mu}g^{\beta\nu}\nabla_\lambda g_{\alpha\beta}=Q_{\lambda}^{~ \mu\nu},
\ee
and using the decomposition of nonmetricity (\ref{Q_dec}) in 4 dimensions,
\be\label{Q_dec4}
Q_{\lambda\mu\nu}=\frac{5Q_\alpha-2\tilde{Q}_\alpha}{18}g_{\mu\nu}+\frac{4\tilde{Q}_{(\mu}g_{\nu)\alpha}-Q_{(\mu}g_{\nu)\alpha}}{9}+\Omega_{\lambda\mu\nu},
\ee
one can simplify (\ref{eq_con}) to
\begin{align}\label{eq_cs}
	0=\frac{4}{9}\left(\tilde{Q}_\lambda+\frac{1}{2}Q_\lambda\right)g^{\mu\nu}+\frac{10}{9}\left(\tilde{Q}^\mu+\frac{1}{2}Q^\mu\right)\delta^\nu_\lambda&\nonumber\\
	-\frac{2}{9}\left(\tilde{Q}^\nu+\frac{1}{2}Q^\nu\right)\delta^\mu_\lambda-\Omega_\lambda^{~\mu\nu}+\zeta^{[\mu\nu]}_{~~~\lambda}.&
\end{align}
Contracting the above equation with $g_{\mu\nu}$ one gets
\be
\frac{8}{3}\left(\tilde{Q}_\lambda+\frac{1}{2}Q_\lambda\right)=0,
\ee
thus we have
\be\label{eq_1/2}
\tilde{Q}_\lambda=-\frac{1}{2}Q_\lambda.
\ee
Using (\ref{eq_1/2}) and contracting (\ref{eq_cs}) with any non-zero anti-symmetric tensor $\Sigma_{\rho[\mu\nu]}$, we find
\be
\zeta^{[\mu\nu]}_{~~~\lambda}=0
\ee
and
\be\label{Omega}
\Omega_\lambda^{~\mu\nu}=0.
\ee
With (\ref{eq_1/2}) and (\ref{Omega}), we obtain a Schr\"{o}dinger connection (\ref{S_Gamma}) as discussed in the previous section.

Furthermore, after a straightforward calculation by inserting \eqref{eq_RR} and \eqref{eq_Qlmn} into \eqref{eq_EinsteinPalatini}, the modified Einstein equation in Palatini formalism can be largely simplified to obtain
\be\label{eq_Einstein}
\mathring{R}_{\mu\nu}-\frac{1}{2}\mathring{R}g_{\mu\nu}=8\pi T_{\mu\nu},
\ee
where $\mathring{R}_{\mu\nu}$ is the Ricci tensor constructed from the Levi-Civita connection $\gamma^\lambda_{\mu\nu}$ which is metric compatible, i.e.,  $\mathring{\nabla}_\alpha g_{\mu\nu}=0$. This means our model in its Palatini formalism is equivalent to the  general relativity.

Now considering the metric formalism in which the connection has to be assumed to depend on the metric in some way a prior. If we adopt the Schr\"{o}ginger connection \eqref{S_Gamma}, then the variation of \eqref{action} w.r.t. $g_{\mu\nu}$ gives the modified Einstein equation in the metric formalism
\bea\label{eq_EinsteinMetric}
\hspace{-0.5cm}\mathring{R}_{\mu\nu}-\frac{1}{2}\mathring{R}g_{\mu\nu}&-&\frac{2}{9}Q_\rho Q^\rho g_{\mu\nu}-\frac{11}{18}Q_\mu Q_\nu+\frac{2}{3}g_{\mu\nu}\nabla_\rho Q^\rho \nonumber\\
\hspace{-0.5cm}&+&\frac{1}{6}g_{\rho\mu}\nabla_\nu Q^\rho+\frac{1}{6}g_{\rho\nu}\nabla_\mu Q^\rho=8\pi T_{\mu\nu}.
\eea

\section{Cosmological applications}\label{sect3}

In the present Section we will consider the cosmological applications of the Weyl-Schr\"{o}dinger gravity theory, as we have introduced it in Section \ref{sect2}.
As a first step in ours study, we will obtain the generalized Friedmann equations of the theory, by assuming a flat, isotropic and homogeneous Universe. We also point out the presence of extra terms, of geometric nature, in the generalized Friedmann equations, which can be interpreted as a dark energy, and which trigger the accelerated expansion of the Universe.  Then, we will reformulate the basic equations in a dimensionless form, and in the redshift space. The existence of a de Sitter type solution will be investigated in detail. Two cosmological models, obtained by obtained various conditions on the dark energy terms, are obtained, and studied in detail. In each case a comparison with the standard $\Lambda$CDM model and a small set of observational data is also performed.

\subsection{Generalized Friedmann equations in Weyl-Schr\"{o}dinger gravity}

We assume first that the Universe
is described by the isotropic, homogeneous and spatially
flat Friedmann–Lemaitre–Robertson–Walker (FLRW) metric, given by
\be\label{FLRW}
\dd s^2=-\dd t^2+a^2(t)\delta_{ij} \dd x^i\dd x^j,
\ee
where $a(t)$ is the scale factor. We also assume that, due to spatial symmetry, the first Weyl vector can be taken to be of the form
\be\label{Qb}
Q_\rho=\left[\omega (t),0,0,0\right].
\ee

Moreover, we consider that the matter content of the Universe can be described as a perfect fluid, characterized by only two thermodynamic parameters, the energy density $\rho$, and the thermodynamic pressure $p$. Hence,  the ordinary matter energy-momentum tensor is given by
\be
T_{\mu\nu}=\rho u_\mu u_\nu+p\left(u_\mu u_\nu +g_{\mu\nu}\right),
\ee
where $u^\mu$ is the normalized four-velocity of the fluid, satisfying the condition $u_\mu u^\mu=-1$. Then, the field equations  (\ref{eq_EinsteinMetric}) give the two generalized Friedmann equations of the Weyl-Schr\"{o}dinger theory as (see Appendix \ref{App_Fri} for their derivation)
\be\label{eq_FLRW1}
\frac{3\dot{a}^2}{a^2}+\frac{2\dot{a}}{a}\omega-\frac{1}{2}\omega ^2+\dot{\omega}=8\pi\rho
\ee
and
\be\label{eq_FLRW2}
-\frac{2\ddot{a}}{a}-\frac{\dot{a}^2}{a^2}-\frac{7\dot{a}}{3a}\omega-\frac{1}{6}\omega ^2-\frac{2}{3}\dot{\omega}=8\pi p.
\ee

By introducing the Hubble function $H$, defined as $H=\dot{a}/a$, we can reformulate the generalized Friedmann equations as
\be
3H^2=8\pi \left(\rho +\rho _{DE}\right)=8\pi \rho_{eff},
\ee
and
\be
2\dot{H}+3H^2=-8\pi \left(p+p_{DE}\right)=-8\pi p_{eff},
\ee
where we have denoted
\be
\rho_{DE}=\frac{1}{8\pi}\left(-\dot{\omega}-2H\omega +\frac{1}{2}\omega ^2\right),
\ee
and
\be
p_{DE}=\frac{1}{8\pi}\left(\frac{2}{3}\dot{\omega}+\frac{1}{6}\omega ^2+\frac{7}{3}H\omega\right),
\ee
respectively. From the generalized Friedman equations we obtain the global energy balance equation, as given by
\be
\dot{\rho}_{eff}+3H\left(\rho_{eff}+p_{eff}\right)=0,
\ee
which can be explicitly written as
\bea\label{cons1}
&&\dot{\rho}+3H(\rho+p)+\frac{1}{8\pi}\frac{d}{dt}\left(-\dot{\omega}-2H\omega +\frac{1}{2}\omega ^2\right)\nonumber\\
&&+\frac{3}{8\pi}H
\left(-\frac{1}{3}\dot{\omega}+\frac{1}{3}H\omega +\frac{2}{3}\omega ^2\right)=0.
\eea

As an indicator of the accelerated/decelerated expansion, we introduce the deceleration parameter $q$, defined as
\be
q=\frac{d}{dt}\frac{1}{H}-1=-\frac{\dot{H}}{H^2}-1.
\ee

With the use of the generalized Friedmann equations we obtain for the deceleration parameter the expression
\be
q=\frac{1}{2}+\frac{3}{2}\frac{p_{eff}}{\rho_{eff}}=\frac{1}{2}+\frac{3}{2}\frac{p+\frac{1}{8\pi}\left(\frac{2}{3}\dot{\omega}+\frac{1}{6}\omega ^2+\frac{7}{3}H\omega\right)}{\rho +\frac{1}{8\pi}\left(-\dot{\omega}-2H\omega +\frac{1}{2}\omega ^2\right)}.
\ee

Once the condition $q<0$ is satisfied, the Universe will enter into an accelerated phase of expansion. Thus a transition can be triggered in the present model by the dynamical evolution of the Weyl field $\omega$.

To simplify the mathematical formalism we introduce a set of dimensionless variables $(\tau, h, r, P,\Omega)$, defined according to the transformations
\be\label{dim}
\tau =H_0t, H=H_0h, \rho=\frac{3H_0^2}{8\pi}r, p=\frac{3H_0^2}{8\pi}P, \omega =H_0 \Omega,
\ee
where $H_0$ is the present-day value of the Hubble function. Then the system of the generalized Friedmann equations takes the following dimensionless form
\be\label{F1}
h^2=r-\frac{2}{3}h\Omega +\frac{1}{6}\Omega ^2-\frac{1}{3}\frac{d\Omega}{d\tau},
\ee
\be\label{F2}
2\frac{dh}{d\tau}+3h^2=-3P-\frac{7}{3}h\Omega-\frac{1}{6}\Omega ^2-\frac{2}{3}\frac{d\Omega }{d\tau}.
\ee

To facilitate the comparison with the observational data we reformulate the cosmological evolution equations in the redshift space, with the redshift variable defined according to
\be
1+z=\frac{1}{a},
\ee
giving
\be
\frac{d}{d\tau}=-(1+z)h(z)\frac{d}{dz}.
\ee

Then in the redshift space the generalized Friedman equations are given by
\be\label{z1}
h^2(z)=r(z)-\frac{2}{3}h(z)\Omega(z) +\frac{1}{6}\Omega ^2 (z)+\frac{1}{3}(1+z)h(z)\frac{d\Omega}{dz},
\ee
\bea\label{z2}
-2(1+z)h(z)\frac{dh(z)}{dz}&+&3h^2 (z)=-3P(z)-\frac{7}{3}h(z)\Omega (z) \nonumber\\
&&-\frac{1}{6}\Omega ^2 (z)+\frac{2}{3}(1+z)h(z)\frac{d\Omega }{dz}.\nonumber\\
\eea

To test the relevance, and the  viability of the cosmological predictions of the Weyl-Schr\"{o}dinger  gravity theory,
 we will perform a detailed comparison of it with the standard $\Lambda
$CDM cosmology, as well as with a small sample of observational data points, obtained for the Hubble function.

In the $\Lambda $CDM model the Hubble function is given by
\begin{equation}
H=H_{0}\sqrt{\frac{\Omega _{m}}{a^{3}}+\Omega _{\Lambda }}=H_{0}\sqrt{%
\Omega _{m}(1+z)^{3}+\Omega _{\Lambda }},
\end{equation}%
where   $\Omega _{m}=\Omega _{b}+\Omega _{DM}$, with $%
\Omega _{b}=\rho _{b}/\rho _{cr}$, $\Omega _{DM}=\rho _{DM}/\rho
_{cr} $ and $\Omega _{\Lambda }=\Lambda /\rho _{cr}$, where $\rho_{cr}$ is the critical density of the Universe. $\Omega _{b}$, $\Omega _{DM}$ and $\Omega _{DE}$  represent the density parameters of the baryonic matter, dark matter, and dark energy,
respectively. The deceleration parameter can be obtained from the relation
\begin{equation}
q(z)=\frac{3(1+z)^{3}\Omega _{m}}{2\left[ \Omega _{\Lambda }+(1+z)^{3}\Omega
_{m}\right] }-1.
\end{equation}

In the following analysis for the matter and dark energy density parameters
of the $\Lambda $CDM model we will use the numerical values $\Omega _{DM}=0.2589$, $\Omega
_{b}=0.0486$, and $\Omega _{\Lambda }=0.6911$, respectively \cite{1g, 1h}. Hence, the
total matter density parameter $\Omega _{m}=\Omega _{DM}+\Omega _{b}=0.3075$, where we have neglected the contribution of the radiation to the total matter energy balance in the late Universe. The present day value of $q$, as predicted by the $\Lambda$CDM model, is thus $q(0)=-0.5912$, indicating that the recent Universe is in an accelerating expansionary stage. For the observational data we use the values of the Hubble functions from the compilation presented in \cite{Bou}. 

\subsection{The de Sitter solution: $h={\rm constant}$}

We look first for exact vacuum solutions of the Weyl-Schr\"{o}dinger cosmological models, with $P=0$, under the condition of a constant expansion rate, with $h=h_0={\rm constant}$. Eq.~(\ref{F2}) then becomes
\be
\frac{d\Omega }{d\tau}+\frac{7}{2}h_0\Omega+\frac{1}{4}\Omega ^2+\frac{9}{2}h_0^2=0,
\ee
with the general solution given by
\be
\Omega (\tau)=h_0 \left\{\sqrt{31} \tanh \left[\frac{1}{4} \sqrt{31} h_0 \left(\tau -4
   c_1\right)\right]-7\right\},
\ee
where $c_1$ is an arbitrary constant of integration. In the limit of large times we have $\lim_{\tau \rightarrow \infty} \Omega {\tau}=\left(\sqrt{31}-7\right)h_0$, that is, the Weyl vector takes negative values when $\tau$ is very large. Then, from Eq.~(\ref{F1}) we obtain the variation of the matter density during the de Sitter type era as
\bea
r(\tau)&=&\frac{1}{4} h_0^2 \Bigg\{12 \sqrt{31} \tanh \left[\frac{1}{4} \sqrt{31} h_0
   \left(\tau -4 c_1\right)\right]\nonumber\\
 &&  +31 \text{sech}^2\left[\frac{1}{4} \sqrt{31} h_0
   \left(\tau -4 c_1\right)\right]-68\Bigg\}.
\eea
In the large time limit the matter density tends to $\lim_{r\rightarrow \infty}r(\tau)=\left(3\sqrt{31}-17\right)h_0^2$, indicating a slight violation of the energy condition $r>0$ at large time intervals.

\subsection{Model I: dark energy models with a linear EOS}

We will consider now dark energy models that do not satisfy anymore the condition of the constancy of the Hubble function. As a first dark energy model in the Weyl-Schr\"{o}dinger gravity theory we assume that the effective pressure and energy density of the dark energy are related by a linear equation of state, given by
\be
p_{DE}(z)=\sigma (z)\rho_{DE}-\frac{\lambda}{8\pi}.
\ee
where $\lambda$ is a constant.  For the parameter $\sigma (z)$ of the dark energy equation of state we adopt the Chevallier-Polarski-Linder (CPL) parametrization \cite{CPL1,CPL2}, so that
\be
\sigma (z)=\sigma _0+\sigma _a \frac{z}{1+z}.
\ee

This form allows to extend the dark energy EOS to very high redshifts, since $\lim_{z\rightarrow \infty}=\sigma _0+\sigma _a$. Hence, the dynamical cosmological evolution equations describing the expansion of the dust Universe, with $P=0$, take the form
\bea\label{M1a}
&&-\frac{2}{3}\left[1+\frac{3}{2}\sigma (z)\right](1+z)h(z)\frac{d\Omega (z)}{dz}+\frac{7}{3}\left[1+\frac{6\sigma (z)}{7}\right]\nonumber\\
&&\times h(z)\Omega (z)+\frac{1}{6}\left[1-3\sigma (z)\right]\Omega ^2(z)+\lambda=0,
\eea
and
\bea\label{M1b}
&&-2(1+z)h(z)\frac{dh(z)}{dz}+3h^2(z)-\lambda \nonumber\\
&&+\sigma (z)\Bigg[-2h(z)\Omega (z)+\frac{1}{2}\Omega ^2(z)+(1+z)h(z)\frac{d\Omega (z)}{dz}\Bigg]\nonumber\\
&&=0,
\eea
respectively.

The system of equations (\ref{M1a}) and (\ref{M1b}) must be integrated with the initial conditions $h(0)=1$, and $\Omega (0)=\Omega _0$.

Once the functions $h(z)$ and $\Omega (z)$ are known as solutions of the evolution equations, the matter density can be obtained as
\be
r(z)=h^2(z)+\frac{2}{3}h(z)\Omega (z)-\frac{1}{6}\Omega ^2(z)-\frac{1}{3}(1+z)h(z)\frac{d\Omega (z)}{dz}.
\ee

\begin{figure*}[htbp]
\includegraphics[scale=0.6]{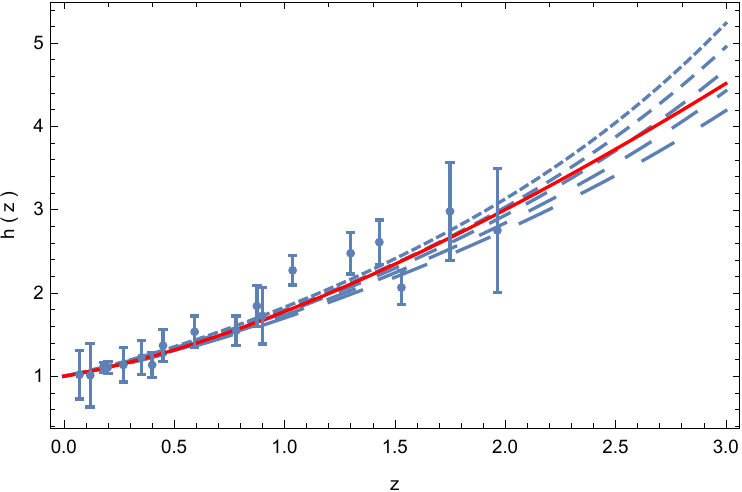}
\includegraphics[scale=0.6]{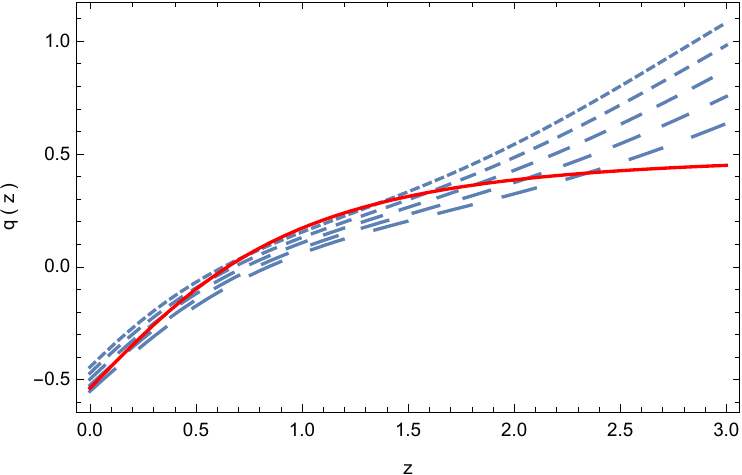}
\caption{Variations as functions of the cosmological redshift of the dimensionless Hubble function $h$ (left panel) and of the deceleration parameter (right) panel in the Weyl-Schr\"{o}dinger cosmological model with a linear equation of state for the dark energy for $\lambda =3.1$ (dotted curve), $\lambda =3.2$ (short dashed curve), $\lambda =3.3$ (dashed curve), $\lambda =3.4$ (long-dashed curve), and $\lambda =3.5$ (ultra-long dashed curve). The initial conditions used to integrate the cosmological evolution equations are $\Omega (0)=-9.7\times 10^{-1}$, and $h(0)=1$, respectively. For the numerical values of the coefficients of the parameter of the dark energy equation of state we have adopted the values $\sigma _0=0.58$ and $\sigma _a=0.0018$, respectively. The observational data are represented with their error bars, while the red curve depicts the predictions of the $\Lambda$CDM model.}\label{fig1}
\end{figure*}

The variations as functions of the redshift of the Hubble function and of the deceleration parameter are represented, for different values of $\lambda$, in Fig.~\ref{fig1}. The Weyl-Schr\"{o}dinger model, closed with an effective equation of state of the dark energy, gives a good description of the observational data, and, for a certain range of the model parameters, can reproduce almost exactly the predictions of the $\Lambda$CDM model. However, some differences do appear in the behavior of the deceleration parameter. Similarly to the standard cosmological models, the Weyl-Schr\"{o}dinger models predicts a decelerating expansion of the Universe at redshifts higher than $z\approx 1$, and an accelerating expansion at lower redshifts.

\begin{figure*}[htbp]
\includegraphics[scale=0.6]{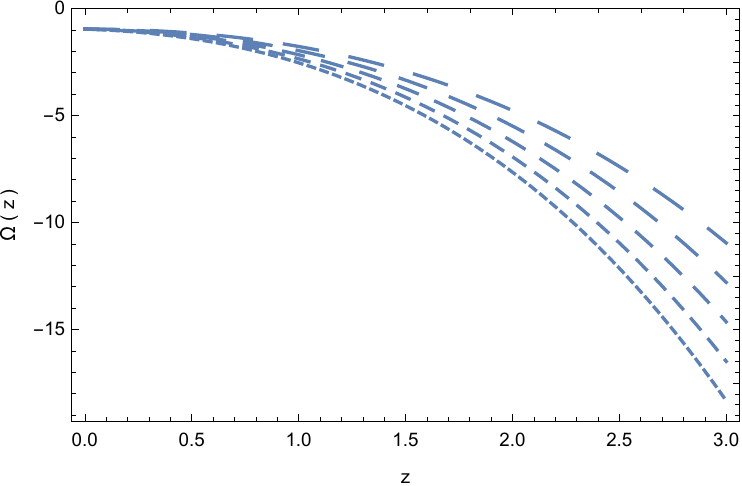}
\includegraphics[scale=0.6]{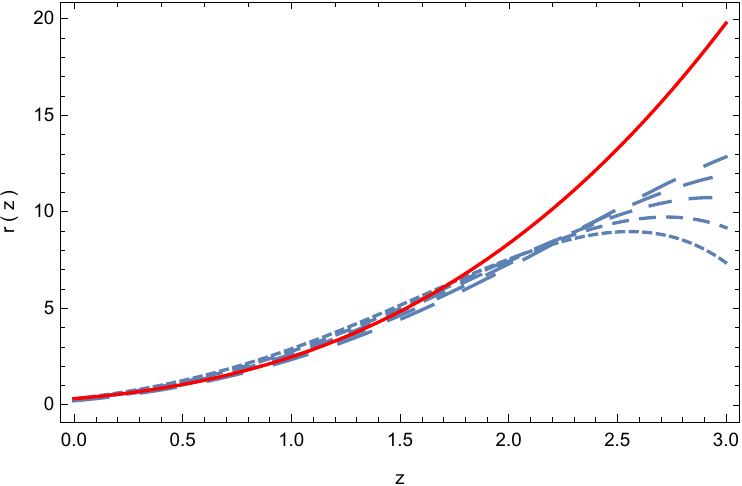}
\caption{Variation as a function of the redshift of the dimensionless Weyl-Schr\"{o}dinger vector  $\Omega$ (left panel) and of the dimensionless matter energy density $r$ in the Weyl-Schr\"{o}dinger model with a linear equation of state for the dark energy  for $\lambda =3.1$ (dotted curve), $\lambda =3.2$ (short dashed curve), $\lambda =3.3$ (dashed curve), $\lambda =3.4$ (long-dashed curve), and $\lambda =3.5$ (ultra-long dashed curve). The initial conditions used to integrate the cosmological evolution equations are $\Omega (0)=-9.7\times 10^{-1}$, and $h(0)=1$, respectively. For the numerical values of the coefficients of the parameter of the dark energy equation of state we have adopted the values $\sigma _0=0.58$ and $\sigma _a=0.0018$, respectively. The red curve depicts the prediction of the $\Lambda$CDM model for the matter energy density, $r(z)=0.3075(1+z)^3$. }\label{fig2}
\end{figure*}

The variations of the temporal component of the Weyl-Schr\"{o}dinger vector $\Omega$, and of the matter energy density $r(z)$ are represented, as a functions of the redshift, in Fig.~\ref{fig2}. The cosmological Weyl-Schr\"{o}dinger vector is a monotonically increasing function of the redshift (a monotonically decreasing function of the cosmological time), and its evolution is strongly dependent, at high redshifts, by the adopted values of the model parameters. Up to a redshift of around $z\approx 0.5$, the cosmological dynamics of the Weyl-Schr\"{o}dinger vector is relatively independent on the numerical values of the model parameter, including the choice of the initial conditions. The matter energy density of the Weyl-Schr\"{o}dinger model coincides, up to a redshift of around $z\approx 2$, with the predictions of the $\Lambda$CDM model. However, at larger redshifts, there are significant differences between the predictions of the two models. Generally, the increase in the matter density occurs faster in the $\Lambda$CDM model, and thus, standard cosmology predicts the existence of a much higher amount of cosmic matter in the early Universe, as compared with the predictions of the Weyl-Schr\"{o}dinger model.

\begin{figure}[htbp]
\includegraphics[scale=0.6]{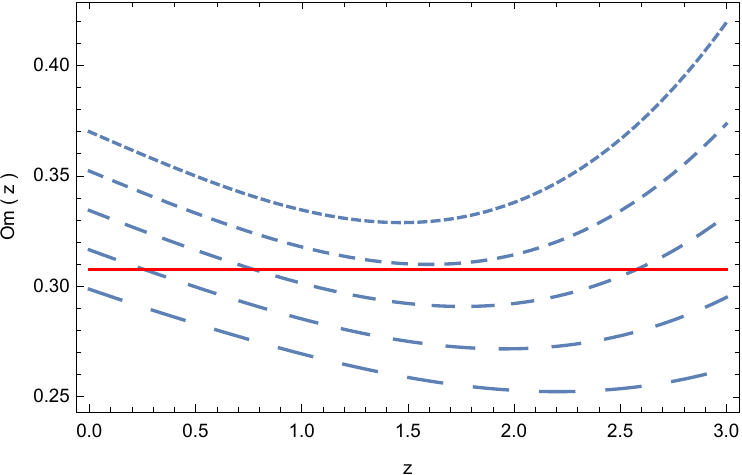}
\caption{Variation of the function $Om(z)$ for the Weyl-Schr\"{o}dinger cosmological model with a linear equation of state of the dark energy for $\lambda =3.1$ (dotted curve), $\lambda =3.2$ (short dashed curve), $\lambda =3.3$ (dashed curve), $\lambda =3.4$ (long-dashed curve), and $\lambda =3.5$ (ultra-long dashed curve). The initial conditions used to integrate the cosmological evolution equations are $\Omega (0)=-9.7\times 10^{-1}$, and $h(0)=1$, respectively.  The red curve corresponds to the prediction of the $\Lambda$CDM model for the $Om(z)$ function. }\label{fig2a}
\end{figure}

Finally, in Fig.~\ref{fig2a}) we present the $Om(z)$ diagnostic of the Weyl-Schr\"{o}dinger cosmological model. The $Om(z)$ diagnostic \cite{Sahni} is
an important theoretical tool which can be used to differentiate alternative cosmological models from the $\Lambda$CDM paradigm. The $Om(z)$ function
is defined as
\be
Om (z)=\frac{H^2(z)/H_0^2-1}{(1+z)^3-1}=\frac{h^2(z)-1}{(1+z)^3-1}.
\ee

In the case of the $\Lambda$CDM model,  $Om(z)$ is a constant, and it is equal to the present day matter density $r(0)=0.3075$. For cosmological models satisfying an equation of state with a constant equation of state parameter $w = {\rm constant}$, the existence of a positive slope of $Om(z)$ is evidence for  a phantom-like evolution, while a negative slope indicates a quintessence-like dynamics. The function $Om)z)$ is represented for the present particular Weyl-Schr\"{o}dinger type cosmological model in Fig.~\ref{fig2a}.

\subsection{Model II: models with conserved matter energy density}

As a second example of a cosmological model in Weyl-Schr\"{o}dinger theory, we consider the case in which both the matter and the Weyl-Schr\"{o}dinger energy-momentum tensors are conserved independently. Hence, we split the total conservation equation (\ref{cons1}) as
\be\label{cons2}
\dot{\rho}+3H(\rho+p)=0,
\ee
and
\bea\label{cons3}
&&\frac{d}{dt}\left(-\dot{\omega}-2H\omega +\frac{1}{2}\omega ^2\right)\nonumber\\
&&+3H\left(-\frac{1}{3}\dot{\omega}+\frac{1}{3}H\omega +\frac{2}{3}\omega ^2\right)=0,
\eea
respectively. For a pressureless dust, Eq.~(\ref{cons2}) can be immediately integrated to give
\be
r(z)=r_0(1+z)^3,
\ee
where $r_0=r(0)$ is the present day matter density. After introducing the dimensionless coordinates as defined in Eqs.~(\ref{dim}), and introducing the new variable $u=d\Omega/d\tau$, the conservation equation of the effective energy of the Weyl-Schr\"{o}dinger field can be reformulated as
\be
-\frac{du}{d\tau}-2\frac{dh}{d\tau}\Omega-3hu+\Omega u+h^2\Omega +2h\Omega ^2=0.
\ee

Hence, in the redfshift space, the cosmological evolution equations of the Weyl-Schr\"{o}dinger theory with conserved matter and vector field effective energy can be formulated as
\be\label{fin1}
(1+z)h(z)\frac{d\Omega}{dz}+u(z)=0,
\ee
\bea\label{fin2}
&&(1+z)h(z)\frac{du(z)}{dz}+2(1+z)h(z)\frac{dh(z)}{dz}\Omega (z)-3h(z) u(z)\nonumber\\
&&+\Omega (z)u(z)+h^2(z)\Omega (z)+2h(z)\Omega ^2(z)=0,
\eea
\bea\label{fin3}
-2(1+z)h(z)\frac{dh(z)}{dz}&+&3h^2 (z)+\frac{7}{3}h(z)\Omega (z) \nonumber\\
&&+\frac{1}{6}\Omega ^2 (z)+\frac{2}{3}u(z)=0.
\eea

The system of differential equations (\ref{fin1})-(\ref{fin3}) must be integrated with the initial conditions $h(0)=1$, $\Omega (0)=\Omega _0$, and $u(0)=u_0$, respectively. However, these initial conditions are not arbitrary, since they must satisfy the constraint, following from the first Friedmann equation (\ref{z1}), which gives
\be\label{u0}
1=r_0-\frac{2}{3}\Omega _0+\frac{1}{6}\Omega _0^2-\frac{1}{3}u_0.
\ee

\begin{figure*}[htbp]
\includegraphics[scale=0.6]{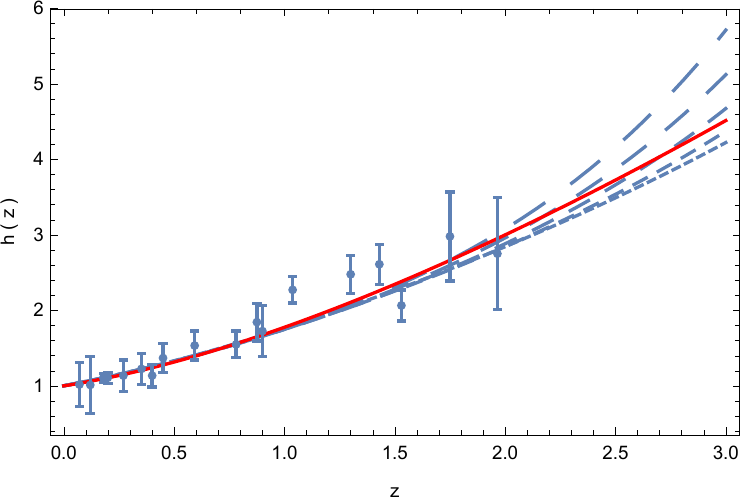}
\includegraphics[scale=0.6]{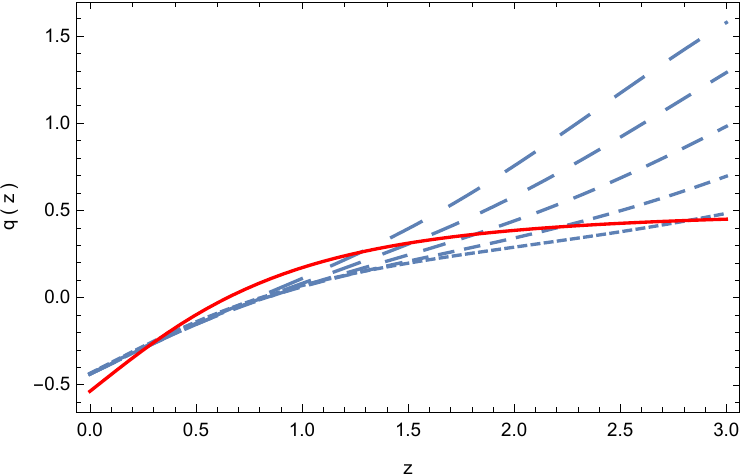}
\caption{Variations as functions of the cosmological redshift of the dimensionless Hubble function $h$ (left panel) and of the deceleration parameter (right) panel in the Weyl-Schr\"{o}dinger cosmological model with conserved matter density, for different values of the initial condition of the Weyl-Schr\"{o}dinger vector field: $\Omega _0 =-0.86$ (dotted curve), $\Omega_0 =-0.90$ (short dashed curve), $\Omega _0 =-0.94$ (dashed curve), $\Omega _0 =-0.98$ (long-dashed curve), and $\Omega _0 =-1.02$ (ultra-long dashed curve). The values of $u_0$ are obtained by using Eq.~(\ref{u0}). The observational data for the Hubble function are represented with their error bars, while the red curve show the theoretical predictions of the $\Lambda$CDM model.}\label{fig3}
\end{figure*}

The variations with respect to the redshift $z$  of the Hubble function and of the deceleration parameter for the Weyl-Schr\"{o}dinger cosmological model with conserved quantities are represented, for different values of $\Omega _0$,  in Fig.~\ref{fig3}. The model gives a good description of the observational data up to a redshift of $z=2$, and can reproduce almost exactly, for specific values of the initial condition $\Omega_0$, $\Omega _0\approx -1$, the predictions of the $\Lambda$CDM model. At redshifts higher than $z=2$, there are some important deviations with respect to the predictions of $\Lambda$CDM. Moreover, in the case of this particular Weyl-Schr\"{o}dinger cosmological model, significant differences do appear in the behavior of the deceleration parameter, which at high redshifts has a very different behavior, as compared with the $\Lambda$CDM predictions.

\begin{figure*}[htbp]
\includegraphics[scale=0.6]{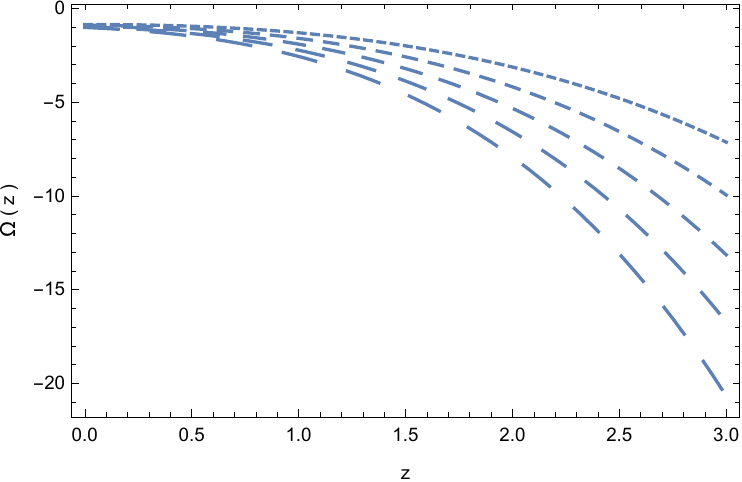}
\includegraphics[scale=0.6]{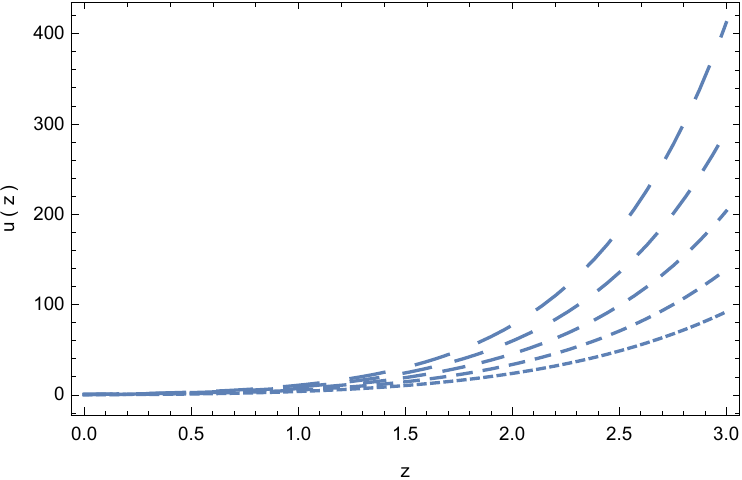}
\caption{Variations as functions of the cosmological redshift of the dimensionless Weyl-Schr\"{o}dinger vector  $\Omega $ (left panel) and of its derivative $u$ (right) panel in the Weyl-Schr\"{o}dinger cosmological model with conserved matter density, for different values of the initial condition of the Weyl-Schr\"{o}dinger vector field: $\Omega _0 =-0.86$ (dotted curve), $\Omega_0 =-0.90$ (short dashed curve), $\Omega _0 =-0.94$ (dashed curve), $\Omega _0 =-0.98$ (long-dashed curve), and $\Omega _0 =-1.02$ (ultra-long dashed curve). The values of $u_0$ are obtained by using Eq.~(\ref{u0}). }\label{fig4}
\end{figure*}

The redshift variations of the Weyl vector $\Omega$, and of its derivative with respect to the redshift $u$ are presented in Fig.~\ref{fig4}.  The Weyl-Schr\"{o}dinger vector field takes negative values, and it is a decreasing function of the redshift. Its behavior at higher redshifts show a strong dependence on the initial condition used to numerically integrate the cosmological evolution equation. The derivative of the Weyl-Schr\"{o}dinger field has only positive values, and it is monotonically increasing function of the redshift. While at low redshifts, in the range $0<z<1.5$, the behavior of $u$ is basically independent on the initial condition for $\Omega _0$, at higher redshifts the behavior of $u$ essentially depends on the initial condition for the Weyl-Schr\"{o}dinger vector field.

\begin{figure}[htbp]
\includegraphics[scale=0.6]{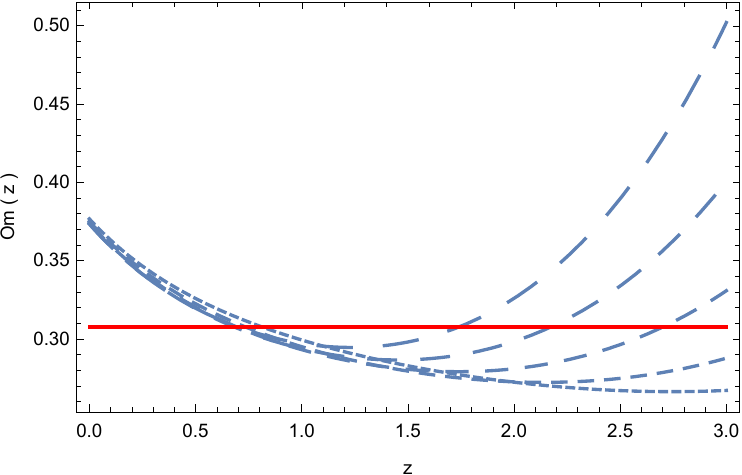}
\caption{Variation of the function $Om(z)$ in the Weyl-Schr\"{o}dinger cosmological model with conserved matter density, for different values of the initial condition of the Weyl-Schr\"{o}dinger vector field: $\Omega _0 =-0.86$ (dotted curve), $\Omega_0 =-0.90$ (short dashed curve), $\Omega _0 =-0.94$ (dashed curve), $\Omega _0 =-0.98$ (long-dashed curve), and $\Omega _0 =-1.02$ (ultra-long dashed curve). The values of $u_0$ are obtained by using Eq.~(\ref{u0}). The prediction of the $\Lambda $CDM model for the $Om(z)$ function is represented by the red solid curve. }\label{fig4a}
\end{figure}

The variation of the $Om(z)$ function for the Weyl-Schr\"{o}dinger cosmological model with conserved matter energy density is represented in Fig.~\ref{fig4a}.

\section{Thermodynamical interpretation of the Weyl-Schr\"{o}dinger theory}\label{sect4}

As one can see from the energy balance equation (\ref{cons1}), in the Weyl-Schr\"{o}dinger gravity theory the matter energy-momentum tensor is not conserved automatically. This aspect also appears in many other modified gravity theories, like, for example, in the  $f\left(R,L_m\right)$ or $f(R,T)$ modified gravity theories with geometry-matter coupling \cite{fRLm, fRT,book}. The interpretation of the nonconservation of the energy of the matter is still a matter of debate, but one of the attractive interpretations of this effect may be related to the possibility of particle creation within the framework of the given theory.

If this is indeed the case, the nonconservation of the energy-momentum tensor of a given gravitational  theory  can be interpreted within the irreversible thermodynamic of open systems. In this framework, we may assume that the non-conservation of the matter energy-momentum tensor in the Weyl-Schr\"{o}dinger gravity theory strongly suggests that, due to the presence of the Weyl geometric effects,  matter generation processes could take place
as a result of the cosmological evolution. The particle creation processes are also a consequence of the quantum field theories in curved space-times, as first shown in
\cite{Parker,Parker1,Parker2}, and they are a direct consequence of the time evolution of the gravitational field. Therefore, the Weyl-Schr\"{o}dinger gravity theory, which also can be interpreted as describing particle creation in a cosmological background, could also lead to the possibility of an effective  semiclassical approach for the description of the quantum field theoretical processes in time-dependent gravitational fields.

\subsection{Brief review of the thermodynamic of irreversible processes in open systems}

In a given physical system, the presence of matter generation processes is related to the basic result that the covariant divergences of
the fundamental equilibrium thermodynamic quantities, like, for example, the energy-momentum tensor, and the particle and entropy fluxes, respectively, are different from zero.
Therefore, all the thermodynamic balance equilibrium equations must be adjusted to incorporate particle creation \cite{P-M,Lima,Su}. For particles that are generated gravitationally, the particle flux
$N^{\mu} \equiv nu^{\mu}$, where $n$ is the particle number density, is described by the balance equation
\begin{equation}\label{n}
\nabla _{\mu}N^{\mu}=\dot{n}+3Hn=n\Psi,
\end{equation}
where by $\Psi $ we have denoted the particle generation rate. If $\Psi \ll H$, the particle creation processes are negligibly small as compared to the expansion rate of the Universe.  We define the entropy flux vector as $S^{\mu} \equiv su^{\mu} = n\sigma u^{\mu}$, where by $s$ we have denoted the entropy density, while $%
\sigma $ represents the entropy per particle. The divergence of the entropy flux must satisfy, according to the second law of thermodynamics, the condition
\begin{equation}  \label{62b}
\nabla _{\mu}S^{\mu}=n\dot{\sigma}+n\sigma \Psi\geq 0.
\end{equation}
By assuming that the entropy density  $\sigma $ is a constant, then
\begin{equation}\label{ent1}
\nabla _{\mu}S^{\mu}=n\sigma \Psi =s\Psi\geq 0.
\end{equation}
Hence, for $\sigma ={\rm constant}$, the variation of the total entropy is exclusively due to the gravitational adiabatic matter creation processes. Because by definition $s>0$, from Eq.~(\ref{ent1}) it follows that the particle creation rate $\Psi$ must satisfy the condition $\Psi \geq 0$. This condition shows that gravitational fields can create matter, but the inverse process is forbidden.

In the presence of particle creation, the energy-momentum tensor of a physical system must also be corrected to include particle creation, as well as the second law of thermodynamics, so that it takes the form \cite{Bar}
\begin{equation}  \label{64}
T^{\mu \nu}=T^{\mu \nu}_\text{eq}+\Delta T^{\mu \nu},
\end{equation}
where $T^{\mu \nu}_\text{eq}$ represents the equilibrium component, while $\Delta T^{\mu \nu}$ is the modification due to the presence of
matter creation. As a consequence of the isotropy and homogeneity of the cosmological space-time, the supplementary  term to the equilibrium energy-momentum tensor describing particle creation must be represented by a scalar quantity. Thus, one can write generally
\begin{equation}
\Delta T_{\; 0}^0=0, \quad \Delta T_{\; i}^j=-p_c\delta_{\; i}^j,
\end{equation}
where $p_c$ is the creation pressure, which represents, in a
phenomenological description, the effects of matter generation in a
macroscopic physical system. Thus, one can represent covariantly the contribution of particle creation to the matter energy-momentum tensor as \cite{Bar}
\begin{equation}
\Delta T^{\mu \nu}=-p_ch^{\mu \nu}=-p_c\left(g^{\mu \nu}+u^{\mu}u^{\nu}\right),
\end{equation}
from which we obtain the condition
\be
u_{\mu}\nabla _{\nu}\Delta T^{\mu \nu}=3Hp_c.
\ee

Hence, in the presence of matter creation, the total thermodynamic energy balance
equation, $u_{\mu}\nabla _{\nu}T^{\mu \nu}=0$,  leads to the generalized energy conservation equation
\begin{equation}\label{cons2i}
\dot{\rho}+3H\left(\rho+p+p_c\right)=0.
\end{equation}

The Gibbs law must also be satisfied by the thermodynamic quantities, and, in the presence of matter creation, it can be
written down as \cite{Lima}
\begin{equation}
n \mathcal{T} \mathrm{d} \left(\frac{s}{n}\right)=n\mathcal{T}\mathrm{d}\sigma=\mathrm{d}\rho -\frac{\rho+p}{n}\mathrm{d}n,
\end{equation}
where by $\mathcal{T}$ we have denoted the thermodynamic temperature of the given system.

\subsection{Thermodynamic quantities in Weyl-Schr\"{o}dinger gravity}

We proceed now to the thermodynamic description of the Weyl-Schr\"{o}dinger gravity. After some simple and straightforward algebraic manipulations, the energy balance equation~(\ref{cons1}) can be rewritten as
\begin{equation} \label{76}
\dot{\rho}+3H\left( \rho +p+p_{c}\right) =0,
\end{equation}%
with the creation pressure $p_{c}$ of the Weyl-Schr\"{o}dinger gravity defined as
\begin{eqnarray}
p_c=\frac{1}{24\pi}\Bigg[-\frac{\ddot{\omega}}{H}+\left(\frac{\omega}{H}-3\right)\dot{\omega}+\left(1-2\frac{\dot{H}}{H^2}\right)H\omega +2\omega ^2\Bigg].\nonumber\\
\end{eqnarray}
 Then, with the use of the creation pressure,  the generalized energy balance equation~(\ref{cons1})
can be obtained, in a way similar to standard general relativistic cosmology, from the vanishing of the divergence of the total energy momentum tensor $%
T^{\mu \nu }$, defined as
\begin{equation}
T^{\mu \nu }=\left( \rho +p+p_{c}\right) u^{\mu }u^{\nu }+\left(
p+p_{c}\right) g^{\mu \nu },
\end{equation}
where, in order to obtain the conservation equation, one must also adopt the comoving frame for the cosmological expansion.

By assuming adiabatic particle production, with $\dot{\sigma}=0$, from the Gibbs law we obtain
\begin{equation}
\dot{\rho}
=\left(\rho+p\right)\frac{\dot{n}}{n}
=\left(\rho+p\right)\left(\Psi-3H\right).
\end{equation}
By using the energy balance equation (\ref{76}) we obtain  the
relation between the creation pressure and the particle creation rate as
\begin{equation}
\Psi=-3H\frac{p_c}{\rho+p}.
\end{equation}

For the Weyl-Schr\"{o}dinger gravity theory for the particle
creation rate we obtain the general expression
\begin{eqnarray}
\Psi&=&-\frac{H}{8\pi \left(\rho +p\right)}\Bigg[-\frac{\ddot{\omega}}{H}+\left(\frac{\omega}{H}-3\right)\dot{\omega}+\left(1-2\frac{\dot{H}}{H^2}\right)H\omega \nonumber\\
&&+2\omega ^2\Bigg].
\end{eqnarray}

The particle creation rate must satisfy the condition $\Psi \geq 0$, which, by taking into account that $H$, $\rho$, and $p$ are all positive, is equivalent to the condition $p_c<0$ for all times. Hence, the condition of the negativity of the creation pressure imposes a strong constraint on the
physical parameters of the Weyl-Schr\"{o}dinger theory. By taking into account that $1-2\dot{H}/H^2=3+2q$, where $q$ is the deceleration parameter, the condition $p_c<0$ can be formulated equivalently as
\bea
(3+2q)\omega <\frac{\ddot{\omega}}{H^2}-\left(\frac{\omega}{H}-3\right)\frac{\dot{\omega}}{H}-2\frac{\omega ^2}{H}.
\eea

The divergence of the entropy flux vector is obtained in terms of
the creation pressure as
\begin{equation}
\nabla _{\mu}S^{\mu}=-3 n \sigma H\frac{ p_c}{\rho +p}.
\end{equation}

The condition $p_c<0$ assures the positivity of the entropy production rate, as required by the second law of thermodynamics. Explicitly, the entropy production rate in the Weyl-Schr\"{o}dinger gravity theory is obtained as
\bea
\nabla _{\mu}S^{\mu}&=&-\frac{ n \sigma H}{8\pi (\rho+p)}\Bigg[-\frac{\ddot{\omega}}{H}+\left(\frac{\omega}{H}-3\right)\dot{\omega}+\left(3+2q\right)H\omega \nonumber\\
&& +2\omega ^2\Bigg].
\eea

We consider now the temperature evolution of the newly created particles in the Weyl-Schr\"{o}dinger gravity.  To obtain the time evolution of a relativistic
fluid in a general framework, we assume that the fluid is described thermodynamically by two equations of state for the density and pressure, which
are given in the general form
\be
\rho =\rho \left(n, \mathcal{T} \right), p=p\left(n,\mathcal{T}\right),
\ee
Then for the time evolution of the matter energy density
we immediately find
\begin{equation}
\dot{\rho}=\left(\frac{\partial \rho }{\partial n} \right)_\mathcal{T}\dot{n}+\left(%
\frac{\partial \rho }{\partial \mathcal{T}} \right)_n\dot{\mathcal{T}}.
\end{equation}

With the use of the energy and particle number balance equations we obtain the relation
\begin{equation}\label{78a}
-3H\left(\rho +p+p_c\right)=\left(\frac{\partial \rho }{\partial n}
\right)_\mathcal{T} n\left(\Psi-3H\right)
+\left(\frac{\partial \rho }{\partial \mathcal{T}} \right)_n\dot{\mathcal{T}}.
\end{equation}
Finally, by using the thermodynamic identity \cite{Bar}
\begin{equation}
\mathcal{T}\left(\frac{\partial p}{\partial \mathcal{T}}\right)_n=\rho+p-n\left(\frac{\partial
\rho}{\partial n}\right)_\mathcal{T},
\end{equation}
from Eq.~(\ref{78a}) we obtain the temperature evolution of the newly created particle in a relativistic fluid as
\begin{equation}\label{102}
\frac{\dot{\mathcal{T}}}{\mathcal{T}}=\left(\frac{\partial p}{\partial \rho}\right)_n\frac{\dot{n}}{n}=c_s^2\frac{\dot{n}}{n},
\end{equation}
where $c_s^2=\left(\partial p/\partial \rho\right)_n$ is the speed of sound. Eq.~(\ref{102}) can be also written as
\be
\frac{\dot{\mathcal{T}}}{\mathcal{T}}=c_s^2\left(\Psi-3H\right)=-3c_s^2H\left(1+\frac{p_c}{\rho+p}\right).
\ee
Hence, in the Weyl-Schr\"{o}dinger gravity theory, the temperature evolution can be obtained in the form
\bea
\frac{\dot{\mathcal{T}}}{\mathcal{T}}&=&-3c_s^2H\Bigg\{1+\frac{1}{24\pi (\rho+p)}\Bigg[-\frac{\ddot{\omega}}{H}+\left(\frac{\omega}{H}-3\right)\dot{\omega}\nonumber\\
&&+\left(3+2q\right)H\omega +2\omega ^2\Bigg]\Bigg\}.
\eea
In order for the temperature of the particles to increase, $\dot{\mathcal{T}}>0$, the thermodynamic condition $1+p_c/(\rho+p)<0$ must be satisfied, which is equivalent, for $\rho >0$, $p>0$,  to $\rho+p+p_c<0$, or $\rho+p<-p_c$.

If $\left(\partial p/\partial \rho\right)_n=c_s^2=\gamma =\mathrm{
constant}>0$, we find the temperature-newly created particle number relation as given by
the simple power law expression $\mathcal{T} \sim n^\gamma$.

\subsubsection{The case of negative particle pressure}

In the present thermodynamical interpretation of the Weyl-Schr\"{o}dinger we have assumed that matter is created in an ordinary form, with positive energy density and pressure, satisfying an equation of state of the form $p=w\rho$. Hence, all our previous results are valid for $w\geq 0$. However,  the considered thermodynamic approach, and the interpretation of the Weyl-Schr\"{o}dinger gravity theory can be also generalized to the case $w<0$, that is, to the case of the creation of exotic particles satisfying an equation of state with a negative parameter $w$. Such particles could be, for example, "dark energy" particles, satisfying the equation of state $\rho+p=0$.  Next, we consider the problem of the negative $w$, and we prove that our interpretation, and results, are still valid, in the sense of well-definiteness and regularity, even in the special case $w=-1$. To investigate this problem more closely we consider the temperature evolution equation
\begin{equation}\label{tempevol}
\frac{\dot{\mathcal{T}}}{\mathcal{T}}
=\left(\frac{\partial p}{\partial \rho}\right)_n\frac{\dot{n}}{n}.
\end{equation}
and we will show that it is still valid even if $w= p/\rho = -1$. The proof of this result is as follows. The perfect-fluid energy-momentum balance equation is given by
\begin{eqnarray}\label{61}
&&\dot{\rho}+3(\rho + p+p_c) H = 0.
\end{eqnarray}
With $w=-1$, Eq.~\eqref{61} becomes
\begin{equation}
  \dot{\rho} \equiv -3H p_c.
\end{equation}
 By assuming again adiabatic particle production, with $\dot{\sigma}=0$, where $\sigma$ is the entropy per particle, from the Gibbs law we obtain
\begin{equation}
\dot{\rho} = (\rho+p)\frac{\dot{n}}{n} = 0.
\end{equation}
From the above two equations it follows that
\begin{equation}
  \dot{\rho} = -3Hp_c = 0, p_c=0.
\end{equation}
On the other hand, from the general equation of state for the density  $\rho = \rho \left(n, \mathcal{T}\right)$, we obtain
\begin{equation}
\dot{\rho}=\left(\frac{\partial \rho }{\partial n}\right)_\mathcal{T} \dot{n} +\left(\frac{\partial \rho}{\partial \mathcal{T}}\right)_n\dot{\mathcal{T}}
= 0.
\end{equation}
From the above equation, and from the thermodynamic identity \cite{Bar},
\begin{equation}
\mathcal{T}\left(\frac{\partial p}{\partial \mathcal{T}}\right)_n
= \rho+p-n\left(\frac{\partial\rho}{\partial n}\right)_\mathcal{T}
= -n\left(\frac{\partial\rho}{\partial n}\right)_\mathcal{T},
\end{equation}
it follows immediately that Eq.~\eqref{tempevol} is still valid even for negative values of $w$, including the value $w=-1$. If $w=-1=\text{constant}$, from Eq.~\eqref{tempevol} we obtain that $n\mathcal{T}={\rm constant}$, or, equivalently,  $\mathcal{T}\sim 1/n$. This interesting relation indicates that the thermodynamic temperature of very low density "dark energy" particles is very high, while systems having a very high particle number density have a very low temperature. For $n\rightarrow \infty$, the temperature of the system of "dark energy" particles tends to zero. Therefore, in the Weyl-Schr\"{o}dinger gravity theory the creation of exotic particles, satisfying linear barotropic equations of state with negative parameter, is also possible.

\section{Discussions and final remarks}\label{sect5}

In the present paper we have considered a gravitational theory based on a geometry that goes beyond the standard Riemannian one. More exactly, we have investigated the physical implications of a geometry proposed a long time ago by Erwin Schr\"{o}dinger \cite{SchroBook}, and which, interestingly enough, despite of its many remarkable features, did not attract much attention in the scientific community. The starting point of Schr\"{o}dinger's  theory is Weyl geometry. In its initial formulation, Weyl, in an attempt to unify the gravitational and the electromagnetic fields, introduced a connection who adds some new terms to the standard Levi-Civita connection of the Riemannian geometry. These extra terms are known generally as the nonmetricity $Q_{\mu \nu \lambda}$. In Weyl's theory under parallel transport not only the direction, but also the length of vectors vary. The trace of the nonmetricity (the Weyl vector) was identified by Weyl as the electromagnetic field potential. However, Einstein severely criticised Weyl's theory, and this criticism led to its long time abandonment \cite{Scholz}. Weyl's theory is based on the principle of conformal invariance,  which has many attractive features, and it is assumed to be a fundamental symmetry of nature \cite{Pen, Pen1, Hooft, Hooft1}, unifying the Standard Model of the elementary particles, and gravitation.

On the other hand, Schr\"{o}dinger  \cite{SchroBook}, tried to overcome Einstein's criticism of the Weyl theory by considering a symmetric connection in which
the length of vectors is not changed under parallel transport, even in the presence of nonmetricity.
The Schr\"{o}dinger connection $\Gamma^\lambda_{\phantom{\lambda}\mu\nu}$ can be defined generally as  \cite{Klemm:2020gfm}
\begin{equation}\label{Schconn}
\Gamma^\lambda_{\phantom{\lambda}\mu\nu} = \gamma^\lambda_{\phantom{\lambda}
\mu\nu} + g^{\lambda\rho} S_{\rho\mu\nu},
\end{equation}
where $S_{\mu\nu\rho}$ is a tensor having the properties
\begin{equation}
S_{\lambda\mu\nu} = S_{\lambda\nu\mu}, \quad S_{(\lambda\mu\nu)} = 0.
\end{equation}
If
\begin{equation}
S_{\lambda\mu\nu} = - Q_{\lambda\mu\nu},
\end{equation}
the length of the vectors is invariant during parallel transport \cite{Klemm:2020gfm}. But, similarly to the standard Riemannian case, the angle between vectors changes due to the parallel transport \cite{SchroBook}.  It is interesting to note that one could consider vanishing nonmetricity, and non-zero torsion,  and then symmetrize the connection in
$\mu,\nu$. Thus, we find
\begin{equation}
{\Gamma^\lambda}_{(\mu\nu)} := \check{\Gamma}^\lambda_{\phantom{\lambda}(\mu\nu)} =\gamma^\lambda_{\phantom{\lambda}\mu\nu} - 2 g^{\lambda\rho} T_{(\mu|\rho|\nu)}.
\end{equation}
Then, if
\begin{equation}
S_{\lambda\mu\nu} = - 2 T_{(\mu|\lambda|\nu)},
\end{equation}
$\check{\Gamma}^\lambda_{\phantom{\lambda}(\mu\nu)}$ is identical  with (\ref{Schconn}) \cite{Klemm:2020gfm}. Hence, it turns out that (\ref{Schconn}) can be written down either with regard to torsion, by using nonmetricity only, or as relating to both nonmetricity and torsion.

In order to formulate the gravitational theory based on the Weyl-Schr\"{o}dinger geometry we have introduced the gravitational action (\ref{action}), which has essentially a very simple mathematical structure. In the absence of torsion,  the action is constructed additively from the Weyl scalar $R$ plus the squares of the two contractions of the nonmetricity $Q_\rho$ and $\tilde{Q}_\rho$, respectively. In order to recover the Schr\"{o}dinger connection one must impose the condition $Q_\rho=-2\tilde{Q}_\rho$, which gives finally the field equations (\ref{eq_EinsteinMetric}), which are the basic equations of the present Weyl-Schr\"{o}dinger theory. The field equations, as well as the corresponding connection, have a very interesting mathematical feature, in the sense that no free arbitrary parameters are introduced in the theory, and all the coefficients in the action, and field equations, are purely numerical. Thus, except the standard gravitational coupling constant of general relativity, no new parameter does appear in the field equations.

In order to consider the physical implications of the Weyl-Schr\"{o}dinger gravity, and its viability, we have analyzed in detail the cosmological models that follow from the theory. As a first step, after adopting the homogeneous, isotropic and flat FLRW metric, and adopting a specific form for the nonmetricity vector, we have obtained the generalized Friedmann equations, in which two new terms do appear. These two terms, representing some extra contributions coming from nonmetricity, can correspond to a dark energy type fluid, whose energy density $\rho _{DE}$ and pressure $p_{DE}$ are completely determined by the temporal component of $Q_\rho$. In the present approach the two generalized Friedmann equations contain four unknowns $(H,\omega, \rho, p)$, and even after imposing an equation of state the system is still over-determined. But this allows to construct various cosmological scenarios, by imposing some physically reasonable conditions on the effective dark energy and pressure. In this context we have considered two distinct cosmological models. In the first model we have imposed a linear equation of state relating the dark energy pressure and density, the equation of state being parameterized by redshift dependent parameter, defined according to the CPL prescription. The model thus obtained, depending on four parameters $\left(\Omega _0,\sigma _0,\sigma _a,\lambda\right)$, can be studied numerically in the redshift space. Once the numerical solution is known, a comparison with a small set of observational data of the Hubble function, and with the $\Lambda$CDM model can be performed. The model describes well the observational data for the Hubble function, and for some specific values of the model parameters the $\Lambda$CDM model can be recovered almost exactly. The matter density as predicted by this Weyl-Schr\"{o}dinger cosmological model coincides with the $\Lambda$CDM predictions up to a redshift of $z\approx 1.5$, but at higher redshifts the predictions of the two models are rather different.

A second simple cosmological model can be obtained by imposing the condition of the conservation of the matter energy density, which is required to satisfy the standard equation $\dot{\rho}+3H\rho=0$. The conservation equation determines the matter energy density as having the same form as in the $\Lambda$CDM model. The cosmological evolution is thus determined by the initial condition $\Omega (0)$ of the Weyl vector, and of its derivative $u (0)$. But the first Friedmann equation gives a constraint at $z=0$, which allows to express $u(0)$ in terms of $r(0)$ and $\Omega (0)$. Hence, in this cosmological model the dynamical evolution depends on a single parameter only, the present day value of the dimensionless Weyl vector $\Omega (0)$.  It is interesting that the value $\Omega (0)=-1$ reproduces (almost) exactly the predictions of the $\Lambda$CDM model for the Hubble function. This gives for the present day value of the temporal component of the Weyl vector $\omega (0)=-H_0$. Hence, this two parameters model, depending on the present day values of the matter density, and with $\omega (0)=-H_0$, represents an intriguing, but effective alternative of the $\Lambda$CDM paradigm.

One of the interesting features of the Weyl-Schr\"{o}dinger cosmology is that generally the matter energy density is not conserved, and an energy transfer from geometry (nonmetricity) to matter may take place, resulting in the production of new particles. The nonconservation of the matter energy-momentum tensor can be interpreted in the framework of the thermodynamics of irreversible processes in open systems. We have developed this interpretation in a systematic way, and we have obtained the particle creation rates, the creation pressure, and the entropy and temperature evolution as a function of the Weyl vector, and of its derivatives.

Several physical mechanisms that allow for the production of particles in gravitational fields are known. Most of these particle creation processes are the result of quantum field theoretical or quantum mechanical effects in curved space-times. Particle production in curved space-times can be briefly described in the following manner (see \cite{Jaume} and \cite{Jaume1}, and references therein). The conformally invariant Lagrangian of a scalar field is $\mathcal{L}=(1/2)\left(\nabla _\mu \nabla ^\mu -m^2\phi^2-\xi R \phi^2\right)$, which gives for the evolution of $\phi$ the generalized Klein-Gordon equation
\begin{equation}
\left(-\nabla _\mu\nabla ^\mu+m^2+\xi R\right)\phi=0.
\end{equation}

From the above Klein-Gordon equation the particle number density $n$ produced by the expansion of the Universe can be obtained, in the adiabatic approximation, as given by $n=mH^2/512 \pi$. The energy density of the created particles is given by $\rho =m^2H^2/96\pi$ \cite{Jaume}. On the other hand,  in the present particle creation model, by considering the zero pressure case, the particle creation rate is given by $\Psi =3H+\dot{\rho}/\rho$, a relation which follows from the energy density balance equation $\dot{\rho}+3H\rho=\Psi \rho$. Substituting the expression of $\Psi$ into the particle balance equation (\ref{n}) it turns out that the newly created matter satisfies the matter density - particle number relation given by $\rho =kn$, where $k$ is a constant. In the simple general relativistic approximation, from the first Friedmann equation we obtain the relation $\rho \propto H^2$, or, equivalently,  $n\propto H^2$. These  results are qualitatively similar to the simple estimations obtained with the use of the quantum field theory in curved space-time. However, in the present Weyl-Schr\"{o}dinger theory, corrections terms to the particle number density, and to the energy density, coming from the nonmetricity of the space-time, are also present.

Matter creation effects could also appear due to vacuum instabilities in gravitational and gauge fields.  These instabilities may be caused  by the conformal trace anomaly
$\Gamma =(\pi /2)\left<T_\mu^\mu\right>$, where $\left<T_\mu^\mu\right>$ is the anomalous trace of the matter energy-momentum tensor $T^{\mu \nu}$ \cite{Chern}.  The relation for $\Gamma$ can describe Schwinger pair creation in massless quantum electrodynamics, the radiation generated by static gravitational fields, or the photon and neutrino pair production \cite{Chern}. Therefore, there are a large number of physical mechanisms that could create particles via quantum mechanisms.  The Weyl-Schr\"{o}dinger theory introduced in the present paper could give, on a classical level, at least some qualitative insights, of the quantum processes that may play an important role in cosmology.

To summarize:  in the present work we have proposed and analyzed in detail, from the point of view of the theoretical consistency, and of the concordance with observations, a
geometrical dark energy model, based on the Weyl-Schr\"{o}dinger theory, which has its origins in the Weyl geometry. In this theory, an effective fluid type dark energy component
can be generated from the non-Riemannian geometric structures that determine the properties of the space-time. The Weyl-Schr\"{o}dinger type model have a  close relationship with the standard general relativistic Friedmann cosmological evolution equations, with the Weyl-Schr\"{o}dinger models exactly reproducing in some particular case the $\Lambda$CDM dynamics. The Weyl-Schr\"{o}dinger  models permit to introduce in a simple and intuitive way a geometric dark energy term, of fluid type, for the description of the
cosmological evolution. The considered Weyl-Schr\"{o}dinger models also give a good description of the cosmological observational data, generally in terms of very few free parameters. They can also (almost) exactly reproduce the predictions of the $\Lambda$CDM standard cosmological model. However, one should emphasize that important differences with standard cosmology do appear at high redshifts, and in the numerical values of some cosmographic quantities. Despite these shortcomings, the Weyl-Schr\"{o}dinger type FLRW cosmological model may become an important and attractive alternative to the $\Lambda$CDM model, in terms of theoretical foundations, explanations of the observational data, and predictive power. These models may also yield some new perspectives, and a better understanding of the intricate relation existing between the physical reality and abstract
mathematical structures.

\section*{Acknowledgments}

The work of T.H. is supported by a grant from
the Romanian Ministry of Education and Research,
CNCS-UEFISCDI, project number PN-III-P4-ID-PCE2020-2255 (PNCDI III). L.M. acknowledges the Project funded by China Postdoctoral Science Foundation (2022M723677). H.H.Z. is supported in part by the National Natural Science Foundation of China (NSFC) under
Grants No. 12275367 and No. 11875327, the Fundamental Research Funds for the
Central Universities, and the Sun Yat-Sen University Science Foundation. S.D.L. is supported by the Natural Science Foundation of Guangdong Province.

\begin{appendix}

\section{Appendix}

In this Appendix we present explicitly the calculational details of the main mathematical results of our approach.
	
\subsection{The variation of the action with respect to $g^{\mu\nu}$ and $\Gamma^\lambda_{~\mu\nu}$ in Palatini formalism}\label{App_var}

Firstly, let us start with the variation of $Q_\rho$ and $\tilde{Q}_\rho$ with respect to $g^{\mu\nu}$. By the definitions
\be
Q_{\rho\mu\nu}\equiv-\nabla_\rho g_{\mu\nu}, Q_\rho\equiv g^{\mu\nu}Q_{\rho\mu\nu},
\ee
 and
 \be
 \tilde{Q}_\rho=g^{\mu\nu}Q_{\mu\nu\rho},
 \ee
 we have
\be
\delta_g Q_{\rho\mu\nu}=-\nabla_\rho\delta g_{\mu\nu}.
\ee

Thus
\begin{align}
	\delta_g Q_\rho&=\delta_g(g^{\mu\nu}Q_{\rho\mu\nu})
	=Q_{\rho\mu\nu}\delta g^{\mu\nu}+g^{\mu\nu}\delta_gQ_{\rho\mu\nu}\nonumber\\
	&=Q_{\rho\mu\nu}\delta g^{\mu\nu}-g^{\alpha\beta}\nabla_\rho\delta g_{\alpha\beta}\nonumber\\
	&=Q_{\rho\mu\nu}\delta g^{\mu\nu}+g^{\alpha\beta}\nabla_\rho(g_{\alpha\mu}g_{\beta\nu}\delta g^{\mu\nu}),
\end{align}
and
\begin{align}
    \delta_g\tilde{Q}_\rho&=\delta_g(g^{\mu\nu}Q_{\mu\nu\rho})
    =Q_{\mu\nu\rho}\delta g^{\mu\nu}+g^{\mu\nu}\delta_g Q_{\mu\nu\rho}\nonumber\\
    &=Q_{\mu\nu\rho}\delta g^{\mu\nu}-g^{\alpha\beta}\nabla_\alpha \delta g_{\beta\rho}\nonumber\\
    &=Q_{\mu\nu\rho}\delta g^{\mu\nu}+g^{\alpha\beta}\nabla_\alpha(g_{\beta\mu}g_{\rho\nu}\delta g^{\mu\nu}),
\end{align}
where $\delta g_{\alpha\beta}=-g_{\alpha\mu}g_{\beta\nu}\delta g^{\mu\nu}$ is used.
With the use of another useful relation, $\delta g=-g g_{\mu\nu}\delta g^{\mu\nu}$, one gets
\be
\delta\sqrt{-g}=-\frac{\delta g}{2\sqrt{-g}}=-\frac{1}{2}\sqrt{-g}g_{\mu\nu}\delta g^{\mu\nu}.
\ee
Then the variation of (\ref{action}) with respect to $g^{\mu\nu}$ gives
\begin{align}\label{deltagS}
	&\delta_g S=\frac{1}{16\pi}\int d^4x \bigg[R\delta\sqrt{-g}+\sqrt{-g}R_{(\mu\nu)}\delta g^{\mu\nu}+\frac{5}{24}\cdot\nonumber\\
	&\bigg(Q_\alpha Q^\alpha\delta\sqrt{-g}+\sqrt{-g}Q_\mu Q_\nu\delta g^{\mu\nu}+2\sqrt{-g}g^{\alpha\beta}Q_\alpha\delta_g Q_\beta\bigg)+\nonumber\\
	&\frac{1}{6}\bigg(\tilde{Q}_\alpha\tilde{Q}^\alpha\delta\sqrt{-g}+\sqrt{-g}\tilde{Q}_\mu\tilde{Q}_\nu\delta g^{\mu\nu}+2\sqrt{-g}g^{\alpha\beta}\tilde{Q}_\alpha\delta_g\tilde{Q}_\beta\bigg)\nonumber\\
	&+2T^\rho Q_\rho\delta\sqrt{-g}+2\sqrt{-g}T^\rho\delta_g Q_\rho-8\pi\sqrt{-g}T_{\mu\nu}\delta g^{\mu\nu}\bigg]\nonumber\\
	&=\frac{1}{16\pi}\int d^4x \bigg\{\sqrt{-g}\delta g^{\mu\nu}\bigg(R_{(\mu\nu)}-\frac{1}{2}Rg_{\mu\nu}\bigg)+\frac{5}{24}\cdot\nonumber\\
	&\bigg[\sqrt{-g}\delta g^{\mu\nu}\bigg(-\frac{1}{2}g_{\mu\nu}Q_\alpha Q^\alpha+Q_\mu Q_\nu+2g^{\alpha\beta}Q_\alpha Q_{\beta\mu\nu}\bigg)+\nonumber\\
	&2\sqrt{-g}g^{\alpha\beta}Q_\alpha g^{\rho\sigma}\nabla_\beta\big(g_{\rho\mu}g_{\sigma\nu}\delta g^{\mu\nu}\big)\bigg]+\frac{1}{6}\cdot\nonumber\\
	&\bigg[\sqrt{-g}\delta g^{\mu\nu}\bigg(-\frac{1}{2}g_{\mu\nu}\tilde{Q}_\alpha\tilde{Q}^\alpha+\tilde{Q}_\mu\tilde{Q}_\nu+2g^{\alpha\beta}\tilde{Q}_\alpha Q_{\mu\nu\beta}\bigg)+\nonumber\\
	&2\sqrt{-g}g^{\alpha\beta}\tilde{Q}_\alpha g^{\rho\sigma}\nabla_\rho(g_{\sigma\mu}g_{\beta\nu}\delta g^{\mu\nu})\bigg]+2\sqrt{-g}T^\rho\cdot\nonumber\\
	&\quad\bigg[\delta g^{\mu\nu}\bigg(Q_{\rho\mu\nu}-\frac{1}{2}Q_\rho g_{\mu\nu}\bigg)+g^{\alpha\beta}\nabla_\rho(g_{\alpha\mu}g_{\beta\nu}\delta g^{\mu\nu})\bigg]\nonumber\\
	&\quad-8\pi\sqrt{-g}T_{\mu\nu}\delta g^{\mu\nu}\bigg\}\nonumber\\
	&=\frac{1}{16\pi}\int d^4x \sqrt{-g}\delta g^{\mu\nu}\bigg\{R_{(\mu\nu)}-\frac{1}{2}Rg_{\mu\nu}+\frac{5}{24}\cdot\nonumber\\
	&\bigg[-\frac{1}{2}g_{\mu\nu}Q_\alpha Q^\alpha+Q_\mu Q_\nu+2g^{\alpha\beta}Q_\alpha Q_{\beta\mu\nu}+4g_{\mu\nu}T_\beta Q^\beta\nonumber\\
	&-\frac{2}{\sqrt{-g}}g_{\rho\mu}g_{\sigma\nu}\nabla_\beta(\sqrt{-g}g^{\alpha\beta}Q_\alpha g^{\rho\sigma})\bigg]+\frac{1}{6}\bigg[-\frac{1}{2}g_{\mu\nu}\tilde{Q}_\alpha\tilde{Q}^\alpha\nonumber\\
	&+\tilde{Q}_\mu\tilde{Q}_\nu+2g^{\alpha\beta}\tilde{Q}_\alpha Q_{\mu\nu\beta}+4T_\mu\tilde{Q}_\nu-\frac{2}{\sqrt{-g}}g_{\sigma\mu}g_{\beta\nu}\cdot\nonumber\\
	&\nabla_\rho(\sqrt{-g}g^{\alpha\beta}\tilde{Q}_\alpha g^{\rho\sigma})\bigg]+2\bigg[T^\rho\bigg(Q_{\rho\mu\nu}-\frac{1}{2}Q_\rho g_{\mu\nu}+2T_\rho g_{\mu\nu}\bigg)\nonumber\\
	&-\frac{1}{\sqrt{-g}}g_{\alpha\mu}g_{\beta\nu}\nabla_\rho(\sqrt{-g}T^\rho g^{\alpha\beta})\bigg]-8\pi T_{\mu\nu}\bigg\}.
\end{align}

Here we note that the total derivative terms of the form
\be
\int d^4x\nabla_\lambda\left(\sqrt{-g}X^\lambda\right)
\ee
for some vector $X^\lambda$ cannot be ignored, but rather instead they result in a net contribution given by~\cite{Burton1999}
\be
2\int d^4x \sqrt{-g}T_\lambda  X^\lambda .
\ee
Due to the presence of an extra term in the covariant derivative of a tensor density, and the generally non-symmetric nature of the connection, we find
\begin{equation}
\begin{aligned}
&\nabla_\lambda\left(\sqrt{-g}X^\lambda\right)\\
=&\partial_\lambda\left(\sqrt{-g}X^\lambda\right)+\Gamma^\lambda_{~\rho\lambda}\sqrt{-g}X^\rho-\Gamma^\rho_{~\rho\lambda}\sqrt{-g}X^\lambda\\
=&\partial_\lambda\left(\sqrt{-g}X^\lambda\right)+\sqrt{-g}(\Gamma^\rho_{~\lambda\rho}-\Gamma^\rho_{~\rho\lambda})X^\lambda\\
=&\partial_\lambda\left(\sqrt{-g}X^\lambda\right)+2\sqrt{-g}T_\lambda X^\lambda.
\end{aligned}
\end{equation}

In our model the torsion does not contribute to the final EoM due to (\ref{0T}), as can be seen below.

Noticing that
\begin{align}\label{nabla1}
&\nabla_\beta(\sqrt{-g}g^{\alpha\beta}Q_\alpha g^{\rho\sigma})\nonumber\\
=&(\nabla_\beta\sqrt{-g})g^{\alpha\beta}Q_\alpha g^{\rho\sigma}+
\sqrt{-g}(\nabla_\beta g^{\alpha\beta})Q_\alpha g^{\rho\sigma}\nonumber\\
&+\sqrt{-g}g^{\alpha\beta}(\nabla_\beta Q_\alpha)g^{\rho\sigma}+\sqrt{-g}g^{\alpha\beta}Q_\alpha\nabla_\beta g^{\rho\sigma}\nonumber\\
=&\sqrt{-g}\bigg(-\frac{1}{2}Q_\alpha Q^\alpha g^{\rho\sigma}+\tilde{Q}^\alpha Q_\alpha g^{\rho\sigma}+g^{\rho\sigma}g^{\alpha\beta}\nabla_\beta Q_\alpha\nonumber\\
&+g^{\alpha\beta}Q_\alpha Q_\beta^{~\rho\sigma}\bigg)
\end{align}
and
\begin{align}\label{nabla2}
	&\nabla_\rho(\sqrt{-g}g^{\alpha\beta}\tilde{Q}_\alpha g^{\rho\sigma})\nonumber\\
	=&(\nabla_\rho\sqrt{-g})g^{\alpha\beta}\tilde{Q}_\alpha g^{\rho\sigma}+
	\sqrt{-g}(\nabla_\rho g^{\alpha\beta})\tilde{Q}_\alpha g^{\rho\sigma}\nonumber\\
	&+\sqrt{-g}g^{\alpha\beta}(\nabla_\rho \tilde{Q}_\alpha)g^{\rho\sigma}+\sqrt{-g}g^{\alpha\beta}\tilde{Q}_\alpha\nabla_\rho g^{\rho\sigma}\nonumber\\
	=&\sqrt{-g}\bigg(-\frac{1}{2}Q_\rho \tilde{Q}_\alpha g^{\alpha\beta}g^{\rho\sigma}+Q_\rho^{~\alpha\beta}\tilde{Q}_\alpha g^{\rho\sigma}+g^{\rho\sigma}g^{\alpha\beta}\nabla_\rho \tilde{Q}_\alpha\nonumber\\
	&+g^{\alpha\beta}\tilde{Q}_\alpha \tilde{Q}^\sigma\bigg),
\end{align}
where (\ref{QL}) and (\ref{QLMN}) are used, we are then able to write down the modified Einstein equation by inserting (\ref{nabla1}) and (\ref{nabla2}) into (\ref{deltagS}) and setting $\delta_g S=0$:
\begin{align}
R_{(\mu\nu)}-\frac{1}{2}Rg_{\mu\nu}+\frac{5}{24}\bigg(\frac{1}{2}g_{\mu\nu}Q_\alpha Q^\alpha
+Q_\mu Q_\nu-2g_{\mu\nu}Q_\alpha \tilde{Q}^\alpha &\nonumber\\ -2g_{\mu\nu}g^{\alpha\beta}\nabla_\beta Q_\alpha\bigg)+\frac{1}{6}\bigg(-\frac{1}{2}g_{\mu\nu}\tilde{Q}_\alpha\tilde{Q}^\alpha-\tilde{Q}_\mu\tilde{Q}_\nu&\nonumber\\
+Q_\mu\tilde{Q}_\nu-2\nabla_\mu\tilde{Q}_\nu\bigg)=8\pi T_{\mu\nu}.&
\end{align}

Now let us turn to the variation of action with respect to $\Gamma^\lambda_{~\mu\nu}$.
One can easily find \cite{Iosifidis:2019dua}
\bea
&&\delta_\Gamma T_\rho=\frac{1}{2}\left(\delta^\mu_\rho\delta^\nu_\lambda-\delta^\mu_\lambda\delta^\nu_\rho\right)\delta\Gamma^\lambda_{\mu\nu},\\
&&\delta_\Gamma R^\alpha_{~\beta\rho\sigma}=\nabla_\rho\delta\Gamma^\alpha_{~\beta\sigma}-\nabla_\sigma\delta\Gamma^\alpha_{~\beta\rho}-2T^\lambda_{~\rho\sigma}\delta\Gamma^\alpha_{~\beta\lambda},\\
&&\delta_\Gamma Q_\rho=2\delta^\nu_\rho\delta^\mu_\lambda\delta\Gamma^\lambda_{\mu\nu},\\
&&\delta_\Gamma\tilde{Q}_\rho=\left(g^{\mu\nu}g_{\rho\lambda}+\delta^\mu_\rho\delta^\nu_\lambda\right)\delta\Gamma^\lambda_{\mu\nu},
\eea
with these the variation of (\ref{action}) leads to
\begin{align}
&\delta_\Gamma S=\int d^4x\sqrt{-g}\bigg[g^{\alpha\beta}\delta_\Gamma R^\rho_{~\alpha\rho\beta}+\frac{5}{12}Q^\rho\delta_\Gamma Q_\rho+\frac{1}{3}\tilde{Q}^\rho\delta_\Gamma \tilde{Q}_\rho\nonumber\\
&\quad\quad+2Q^\rho\delta_\Gamma T_\rho+2T^\rho\delta_\Gamma Q_\rho+\zeta^{[\mu\nu]}_{~~\lambda}\delta\Gamma^\lambda_{~\mu\nu}\bigg]\nonumber\\
&=\int d^4x\sqrt{-g}\bigg[g^{\alpha\beta}\bigg(\nabla_\rho\delta\Gamma^\rho_{~\alpha\beta}-\nabla_\beta\delta\Gamma^\rho_{~\alpha\rho}-2T^\lambda_{~\rho\beta}\delta\Gamma^\rho_{~\alpha\lambda}\bigg)\nonumber\\
&\quad\quad+\delta\Gamma^\lambda_{~\mu\nu}\cdot\bigg(\frac{5}{6}Q^\nu\delta^\mu_\lambda+\frac{1}{3}\tilde{Q}_\lambda g^{\mu\nu}+\frac{1}{3}\tilde{Q}^\mu\delta^\nu_\lambda\nonumber\\
&\quad\quad+Q^\mu\delta^\nu_\lambda-Q^\nu\delta^\mu_\lambda+4T^\nu\delta^\mu_\lambda+\zeta^{[\mu\nu]}_{~~~\lambda}\bigg)\bigg]\nonumber\\
&=\int d^4x\delta\Gamma^\lambda_{~\mu\nu}\bigg[-\nabla_\lambda(\sqrt{-g}g^{\mu\nu})+\nabla_\beta(\sqrt{-g}g^{\mu\beta})\delta^\nu_\lambda+\nonumber\\
&\quad\quad2\sqrt{-g}\bigg(T_\lambda g^{\mu\nu}-T^\mu\delta^\nu_\lambda-g^{\mu\beta}T^\nu_{~\lambda\beta}+2T^\nu\delta^\mu_\lambda\bigg)+\nonumber\\
&\quad\sqrt{-g}\bigg(
\frac{1}{3}\tilde{Q}_\lambda g^{\mu\nu}+\frac{1}{3}\tilde{Q}^\mu\delta^\nu_\lambda+Q^\mu\delta^\nu_\lambda-\frac{1}{6}Q^\nu\delta^\mu_\lambda+\zeta^{[\mu\nu]}_{~~~\lambda}\bigg)\bigg],
\end{align}
and thus $\delta_\Gamma S=0$ gives
\begin{align}
	-\frac{\nabla_\lambda\left(\sqrt{-g}g^{\mu\nu}\right)}{\sqrt{-g}}+\frac{\nabla_\rho\left(\sqrt{-g}g^{\mu\rho}\right)\delta^\nu_\lambda}{\sqrt{-g}}+\frac{1}{3}\tilde{Q}_\lambda g^{\mu\nu}&\nonumber\\
	+\frac{1}{3}\tilde{Q}^\mu\delta^\nu_\lambda+Q^\mu\delta^\nu_\lambda-\frac{1}{6}Q^\nu\delta^\mu_\lambda+\zeta^{[\mu\nu]}_{~~~\lambda}=0.&
\end{align}
Note that again the torsion terms disappear in the EoM.

\subsection{The variation of the action with respect to $g^{\mu\nu}$ in metric formalism}\label{varg}

Since in metric formalism the connection is assumed to depend on the metric in the way given by \eqref{S_Gamma}, the variation with respect to $g_{\mu\nu}$ now has the extra contribution $\delta_g R_{\mu\nu}$, as compared to \eqref{deltagS}. To calculate this contribution, note that
\be
\delta_g Q^\alpha_{~\mu\nu}=-\delta_g\left(g^{\alpha\lambda}\nabla_\lambda g_{\mu\nu}\right)=-g^{\alpha\lambda}\nabla_\lambda \delta g_{\mu\nu}-\delta g^{\alpha\lambda}\nabla_\lambda g_{\mu\nu}
\ee
and
\be
\delta_g\tilde{Q}_\rho=-\frac{1}{2}\delta_g Q_\rho=-\frac{1}{2}Q_{\rho\mu\nu}\delta g^{\mu\nu}+\frac{1}{2}g^{\alpha\beta}\nabla_\rho\delta g_{\alpha\beta},
\ee
then from \eqref{eq_RR}we have
\begin{align}\label{deltaR}
	&\int d^4x\sqrt{-g}g^{\mu\nu}\delta_g R_{\mu\nu}\nonumber\\
	=&\int d^4x\sqrt{-g}g^{\mu\nu}\left(\delta_g\mathring{R}_{\mu\nu}-\mathring{\nabla}_\alpha\delta_g Q^\alpha_{~\mu\nu}+\mathring{\nabla}_\nu\delta_g\tilde{Q}_\mu\right.\nonumber\\
	&\left.+Q^\rho_{~\mu\nu}\delta_g\tilde{Q}_\rho+\tilde{Q}_\rho\delta_g Q^\rho_{~\mu\nu}-Q^\rho_{~\mu\alpha}\delta_g Q^\alpha_{~\rho\nu}-Q^\alpha_{~\rho\nu}\delta_g Q^\rho_{~\mu\alpha}\right)\nonumber\\
	=&\int d^4x\sqrt{-g}g^{\mu\nu}\left[Q^\rho_{~\mu\nu}\left(-\frac{1}{2}Q_{\rho\alpha\beta}\delta g^{\alpha\beta}+\frac{1}{2}g^{\alpha\beta}\nabla_\rho\delta g_{\alpha\beta}\right)\right.\nonumber\\
	&\quad\quad-\frac{1}{2}Q_\rho\left(-g^{\rho\lambda}\nabla_\lambda\delta g_{\mu\nu}-\delta g^{\rho\lambda}\nabla_\lambda g_{\mu\nu}\right)\nonumber\\
	&\quad\quad-Q^\rho_{\mu\alpha}\left(-g^{\alpha\lambda}\nabla_\lambda\delta g_{\rho\nu}-\delta g^{\alpha\lambda}\nabla_\lambda g_{\rho\nu}\right)\nonumber\\
	&\quad\quad-Q^\alpha_{\rho\nu}\left(-g^{\rho\lambda}\nabla_\lambda\delta g_{\mu\alpha}-\delta g^{\rho\lambda}\nabla_\lambda g_{\mu\alpha}\right)\bigg]\nonumber\\
	=&\int d^4x\sqrt{-g}\left(-\frac{1}{2}Q^\rho Q_{\rho\alpha\beta}\delta g^{\alpha\beta}+\frac{1}{2}Q^\rho g^{\alpha\beta}\nabla_\rho\delta g_{\alpha\beta}\right.\nonumber\\
	&\quad\quad+\frac{1}{2}g^{\mu\nu}Q^\lambda\nabla_\lambda\delta g_{\mu\nu}-\frac{1}{2}g^{\mu\nu}Q_\rho\delta g^{\rho\lambda}Q_{\lambda\mu\nu}\nonumber\\
	&\quad\quad+Q^{\rho\nu\lambda}\nabla_\lambda\delta g_{\rho\nu}-Q^{\rho\nu}_{~~\alpha}\delta g^{\alpha\lambda}Q_{\lambda\rho\nu}\nonumber\\
	&\quad\quad+Q^{\alpha\lambda\mu}\nabla_\lambda\delta g_{\mu\alpha}-Q^{\alpha~\mu}_{~\rho}\delta g^{\rho\lambda}Q_{\lambda\mu\alpha}\bigg)\nonumber\\
	=&\int d^4x\left[\sqrt{-g}\delta g^{\mu\nu}\left(-\frac{1}{2}	Q^\rho Q_{\rho\mu\nu}-\frac{1}{2}Q_\mu Q_\nu\right.\right.\nonumber\\
	&\quad\quad-Q^{\rho\sigma}_{~~\mu}Q	_{\nu\rho\sigma}-Q^{\rho~\sigma}_{~\mu}Q_{\nu\sigma\rho}\bigg)\nonumber\\
	&-\frac{1}{2}\delta g_{\alpha\beta}\nabla_\rho\left(\sqrt{-g}Q^\rho g^{\alpha\beta}\right)-\frac{1}{2}\delta g_{\mu\nu}\nabla_\lambda\left(\sqrt{-g}Q^\lambda g^{\mu\nu}\right)\nonumber\\
	&-\delta g_{\rho\nu}\nabla_\lambda\left(\sqrt{-g}Q^{\rho\nu\lambda}\right)-\delta g_{\mu\alpha}\nabla_\lambda\left(\sqrt{-g}Q^{\alpha\lambda\mu}\right)\bigg]\nonumber\\
	=&\int d^4x \sqrt{-g}\delta g^{\mu\nu}\bigg[-\frac{1}{2}Q^\rho Q_{\rho\mu\nu}-\frac{1}{2}Q_\mu Q_\nu-2Q^{\rho\sigma}_{~~\mu}Q_{\nu\rho\sigma}\nonumber\\
	&+\frac{1}{\sqrt{-g}}g_{\alpha\mu}g_{\beta\nu}\bigg(\nabla_\rho(\sqrt{-g}Q^\rho g^{\alpha\beta})+2\nabla_\lambda(\sqrt{-g}Q^{\alpha\beta\lambda})\bigg)\bigg]\nonumber\\
	=&\int d^4x\sqrt{-g}\delta g^{\mu\nu}\bigg(\frac{1}{2}Q^\rho Q_{\rho\mu\nu}-\frac{1}{2}Q_\mu Q_\nu-2Q^{\rho\sigma}_{~~\mu}Q_{\nu\rho\sigma}\nonumber\\
	&-\frac{1}{2}Q_\rho Q^\rho g_{\mu\nu}+g_{\mu\nu}\nabla_\rho Q^\rho-Q_\lambda Q_{\mu\nu}^{~~\lambda}+2g_{\alpha\mu}g_{\beta\nu}\nabla_\lambda Q^{\alpha\beta\lambda}\bigg).
\end{align}

With the solution (\ref{eq_1/2}), we are able to write (\ref{Q_dec4}) as
\be\label{eq_Qlmn}
Q_{\lambda\mu\nu}=\frac{1}{3}Q_\lambda g_{\mu\nu}-\frac{1}{6}Q_\mu g_{\nu\lambda}-\frac{1}{6}Q_\nu g_{\lambda\mu},
\ee
and noticing that because of $Q_{(\rho\mu\nu)}=0$,
\be
\delta g^{\mu\nu}Q_\lambda Q_{\mu\nu}^{~~\lambda}=\delta g^{\mu\nu}Q^\rho\frac{Q_{\mu\nu\rho}+Q_{\nu\mu\rho}}{2}=-\frac{1}{2}\delta g^{\mu\nu}Q^\rho Q_{\rho\mu\nu}
\ee
and
\begin{align}
	2\delta g^{\mu\nu} g_{\alpha\mu}g_{\beta\nu}\nabla_\lambda Q^{\alpha\beta\lambda}=&2\delta g^{\mu\nu}g_{\alpha\mu}g_{\beta\nu}\nabla_\lambda\frac{Q^{\alpha\beta\lambda}+Q^{\beta\alpha\lambda}}{2}\nonumber\\
	=&-\delta g^{\mu\nu} g_{\alpha\mu}g_{\beta\nu}\nabla_\lambda Q^{\lambda\alpha\beta},
\end{align}
\eqref{deltaR} can be further simplified to
\begin{align}
	&\int d^4x\sqrt{-g}g^{\mu\nu}\delta_g R_{\mu\nu}\nonumber\\
	=&\int d^4x \sqrt{-g}\delta g^{\mu\nu}\bigg(-\frac{2}{9}g_{\mu\nu}Q^\rho Q_\rho-\frac{11}{18}Q_\mu Q_\nu\nonumber\\
	&+\frac{2}{3}g_{\mu\nu}\nabla_\rho Q^\rho+\frac{1}{6}g_{\rho\mu}\nabla_\nu Q^\rho+\frac{1}{6}g_{\rho\nu}\nabla_\mu Q^\rho\bigg).
\end{align}
All the other contributions to the Einstein equation will be the same as in the Palatini case. Hence, the field equation in the metric formalism can be obtained by simply adding the above terms into \eqref{eq_Einstein}, and thus we finally arrive at \eqref{eq_EinsteinMetric}.

\subsection{Calculating the Friedmann equations}\label{App_Fri}

According to (\ref{FLRW}) and (\ref{Qb}), all the non-zero components of $Q_{\lambda\mu\nu}$ are
\be
Q_{0ii}=\frac{1}{3}a^2b \quad \text{and} \quad Q_{ii0}=Q_{i0i}=-\frac{1}{6}a^2b,
\ee	
where $i=1,2$ or $3$.
Note that $\gamma^\lambda_{\mu\nu}$ has the same structure,
\be
\gamma^0_{ii}=a\dot{a} \quad \text{and} \quad \gamma^i_{i0}=\gamma^i_{0i}=\frac{\dot{a}}{a},
\ee
we can then calculate all the non-zero components of the Schr\"{o}dinger connection (\ref{S_Gamma}),
\be
\Gamma^0_{~ii}=a\dot{a}+\frac{1}{3}a^2b \quad \text{and} \quad \Gamma^i_{~i0}=\Gamma^i_{~0i}=\frac{\dot{a}}{a}+\frac{1}{6}b,
\ee	
and get the well known results for $\mathring{R}_{\mu\nu}$
\be
\mathring{R}_{00}=3\frac{\dot{a}^2}{a^2} \quad \text{and} \quad \mathring{R}_{ii}=-2\frac{\ddot{a}}{a}-\frac{\dot{a}^2}{a^2}.
\ee

On the other hand, since
\be
Q_0=-2\tilde{Q}_0=b \quad \text{and} \quad Q_i=\tilde{Q}_i=0,
\ee
one gets
\be	
Q_\alpha Q^\alpha=-b^2,	
\ee
\be
\nabla_0 Q^0=\partial_0 Q^0=-\dot{b},\quad \nabla_i Q^i=\Gamma^i_{~0i}Q^0=-\frac{\dot{a}}{a}b-\frac{1}{6}b^2
\ee
and
\be	
\nabla_\rho Q^\rho=-\dot{b}-3\frac{\dot{a}}{a}b-\frac{1}{2}b^2.
\ee

Inserting all the above results into (\ref{eq_EinsteinMetric}), and considering $T_{00}=\rho$ and $T_{ii}=p$, we are able to write down the Friedmann equations as (\ref{eq_FLRW1}) and (\ref{eq_FLRW2}).

\end{appendix}

\end{document}